\documentclass[12pt]{article}

\usepackage{graphicx}
\usepackage{natbib}
\usepackage{hyperref}
\usepackage{amssymb}
\usepackage{amsmath}
\usepackage{physics}
\usepackage{tikz}
\usepackage{pgfplots}
\pgfplotsset{compat = newest}
\usetikzlibrary{decorations.pathreplacing,calligraphy,calc}
\usetikzlibrary{automata, positioning}
\usepackage{nicematrix}
\NiceMatrixOptions{
code-for-first-row = \color{black} ,
code-for-first-col = \color{black} ,
}

\usepackage{multirow}
\usepackage{booktabs}

\def\spacingset#1{\renewcommand{\baselinestretch}%
{#1}\small\normalsize} \spacingset{1}

\usepackage{algorithm}
\usepackage{algpseudocode} % or algorithmicx with algpseudocode
\usepackage{mathtools}

\begin{document}

\def\spacingset#1{\renewcommand{\baselinestretch}%
{#1}\small\normalsize} \spacingset{1}

%%%%%%%%%%%%%%%%%%%%%%%%%%%%%%%%%%%%%%%%%%%%%%%%%%%%%%%%%%%%%%%%%%%%%%%%%%%%%%
\begin{center}
{\LARGE\bf Conflict Forecasting via Conformal Prediction for Markov Processes}
\end{center}

% \author[1]{Aditya Basarkar}

% \author[2]{Emmett B. Kendall}

% \author[3]{David Randahl}

% \author[1]{Jonathan P. Williams}

% \author[4]{Gudmund H. Hermansen}

% \authormark{Basarkar \textsc{et al.}}
% \titlemark{Conflict Forecasting through Conformal Prediction with Hidden Markov Models}

% \address[1]{\orgdiv{Department of Statistics}, \orgname{North Carolina State University}, \orgaddress{\state{North Carolina}, \country{USA}}}

% \address[2]{\orgdiv{Department of Mathematical Sciences}, \orgname{The University of Texas at Dallas}, \orgaddress{\state{Texas}, \country{USA}}}

% \address[3]{\orgdiv{Department of War Studies}, \orgname{Swedish Defense University}, \orgaddress{\state{Stockholm}, \country{Sweden}}}

% \address[4]{\orgdiv{Department of Mathematics}, \orgname{University of Oslo}, \orgaddress{\state{Oslo}, \country{Norway}}}

% \corres{Emmett B. Kendall, \email{emmett.kendall@utdallas.edu}}

% \presentaddress{Mathematical Sciences (FO35), 800 W. Campbell Rd.
% Richardson, TX 75080-3021, USA}

\begin{center}
Aditya Basarkar$^1$, Emmett B. Kendall$^2$, David Randahl$^3$,\\ Jonathan P. Williams$^1$, and Gudmund H. Hermansen$^4$
\end{center}

\begin{center}
\small
$^1$Department of Statistics, North Carolina State University\\
$^2$Department of Mathematical Sciences, The University of Texas at Dallas\\
$^3$Department of War Studies, Swedish Defense University\\
$^4$Department of Mathematics, University of Oslo
\end{center}

\bigskip
\begin{abstract}
Whether or not a country is at war, or experiencing escalating or deescalating levels of conflict, has massive ramifications on a country's national and foreign policy. Given a country's history of conflict, or lack thereof, future predictions about the war-status of a country are valuable information. In this paper, we present the use of conformal prediction on temporally-dependent data to obtain prediction sets of possible future conflict state-sequences. More specifically, we compare the results of conformal prediction to a likelihood-based prediction strategy when the data are assumed to come from a discrete-state Markov process. A point-prediction may not supply sufficient information because the penalty for a wrong prediction is extreme, and so we consider a machine learning alternative that gives valid uncertainty quantification and is robust to model misspecification.  In the data analysis, we present real forecasts of conflict dynamics across multiple countries. Lastly, we assess the differences between the conformal and likelihood-based prediction sets applied to training data which fall under a pervasive edge-case, not discussed in existing literature.
\end{abstract}

\noindent%
{\it Keywords:} conflict dynamics, multistate model, non-exchangeable data, state-space model, time series
\vfill

\newpage
\spacingset{1} % DON'T change the spacing!

\section{Motivation}\label{sec:motive}

The last decade has seen a surge of interest in forecasting conflict dynamics within peace and conflict research. Driven by increased access to fine-grained data and advances in machine learning and computing, several large-scale forecasting projects have emerged targeting armed conflict and related phenomena such as coups, electoral violence, and regime change \citep[e.g.,][]{hegre2019views, hegre2021views2020,randahl2025forecasting,ward2017lessons, morgan2019varieties}. Early efforts focused on point predictions, but there is growing recognition that uncertainty quantification is equally critical: incorrect predictions carry severe consequences, and decision-makers require not just a single forecast but a calibrated sense of the range of plausible outcomes \citep{hegre20252023,randahl2026bin}. This has motivated a shift toward probabilistic forecasting methods and prediction intervals/sets.

Among the models that have shown promise for conflict dynamics are Markov models, which represent the evolution of a conflict system as a sequence of states transitioning over time \citep{randahl2022predicting, williams2024bayesian}. Combined with estimated conditional distributions of fatality counts given the current state, these models can generate forecasts of future conflict trajectories \citep{randahl2024forecasting, hegre20252023}. However, their uncertainty estimates are sensitive to model misspecification, a serious concern given the difficulty of correctly characterizing conflict processes.

Conformal prediction (CP) offers a principled alternative. By requiring only an exchangeability assumption rather than a correctly specified model, CP produces prediction sets with finite-sample marginal coverage guarantees. In this paper, we explore applying CP to discrete-state Markov processes to generate valid prediction sets for future conflict state-sequences, and we compare its performance against likelihood-based approaches. A key challenge we address is that Markov processes, by definition, violate the standard exchangeability assumption underlying conformal methods, and we discuss the implications and limitations of existing approaches for handling their temporal dependence.

The remainder of this article is structured as follows. Section \ref{sec:statback} provides background on the statistical strategies being employed, followed by Section \ref{sec:method} which provides an in-depth description of our methodology and algorithm. Next, Section \ref{sec:experiments} presents the results of both the simulation study and real data analysis. Then, Section \ref{sec:limitations} investigates the performance of these predictions methods for an edge-case that is particularly relevant to our real data. Lastly, Section \ref{sec:conclusion} shares final remarks and conclusions.

\section{Background}\label{sec:statback}
CP offers a robust approach to construct prediction sets that obtain valid coverage for exchangeable data \citep[see][for more on CP]{shafer2008tutorial,vovk2022}. The development of CP has proven to be a very powerful statistical strategy given its finite-sample validity guarantees. Its utility is elucidated by the breadth of fields applying CP methods. These examples include real estate \citep{hjort2024clustered, hjort2025uncertainty}, autonomous cars \citep{bang2024safe}, spatial statistics \citep{mao2024valid}, networks \citep{dashevskiy2008network, luo2023anomalous}, and more. However, CP relies on the assumption that the data are exchangeable, which in many applications, need not be the case. Hence, much of the aforementioned research, as well as what is presented here, discusses how to reformulate the CP algorithm to maintain its statistical guarantees when the data are not exchangeable. The focus for this work is on time series data, particularly Markov processes, where sequential observations are temporally dependent.

Methodological developments in CP for time series data has a well-studied literature \citep[e.g.,][]{xu2021conformal, xu2023conformal,zheng2024conformal, dua2025conformal, neglia2025rescp}. Many of these approaches focus on developing provably valid, or approximately valid, methods for data that exhibit dependencies which violate the exchangeability assumption. One example is the development of adaptive CP, which calibrates the guaranteed confidence level in an online setting to account for on-going distributional changes in the data \citep{gibbs2021adaptive, zaffran2022adaptive, su2024adaptive}. Other approaches seek to re-frame \textit{where} the notion of exchangeability is applied.

This latter approach can be split into two different frameworks, specifically for handling discrete-state Markov processes. The first idea is to treat each state-sequence as an individual observation, and then multiple observations of these state-sequences are assumed to be independent and thus exchangeable \citep{cherubin2016, stankeviciute2021conformal}. However, instead of relying on multiple realizations of independent state-sequences, the second idea leverages properties of Markov processes and permutation-groups to arrive at a notion of partial-exchangeability for a \textit{single} state-sequence \citep{chernozhukov2018exact, chernozhukov2021exact, nettasinghe2023extending} (more detail in Section \ref{subsec:cp}). In either case, the output of the CP algorithm remains a prediction set of forecasted state-sequences. 

Further research has been conducted to extend CP to doubly stochastic processes, like via CP for hidden Markov models (HMMs) \citep{cherubin2016, nettasinghe2023extending, su2024adaptive}. For the purposes of our work, particular focus is afforded to the methodology developed in \cite{nettasinghe2023extending} because, as will be discussed in our real data application, our focus is to forecast CP sets for individual state-sequences which requires the use of permutation-groups and partial-exchangeability. While the algorithmic procedure developed in \cite{nettasinghe2023extending} is for discrete-state HMMs, we apply it to the case of discrete-state Markov models.

Our contributions to the literature are as follows. First, applying this CP framework for Markov processes within the conflict research domain is a new application in itself. Second, we study the robustness of the CP approach proposed in \cite{nettasinghe2023extending} by investigating the efficiency of the forecasted CP sets when the historical training data exhibits both high \textit{and} low levels of variation with respect to state-transitions. As motivated by features of the real data, we explore the efficacy of forecasting state-sequences from a \textit{constant} history (i.e., only one unique state is observed in the training data), which is not considered in \cite{nettasinghe2023extending}. Hence, our commentary and discovery of an edge-case for this partial-exchangeability CP method serves as an additional contribution.

\section{Methodology}\label{sec:method}
Let $\{X_t\}_{t=1}^T$ denote a discrete-state Markov process observed over $T > 0$ time points, where $X_t \in \mathcal{X} = \{1,2,\hdots, m\}$, $m \in \mathbb{Z}^+$. We refer to $\{X_t\}_{t=1}^T$ as the \textit{calibration} sequence/data. The ultimate goal for this work is to forecast state-sequences $T_1 > 0$ time points into the future. Rather than constructing a point-prediction (i.e., one future state-sequence of length $T_1$), our goal is to construct a prediction set, for any level $1-\alpha \in [0,1]$, of state-sequences, say $\mathcal{C}_{1-\alpha}$, that guarantees the true underlying state-sequence is an element of the set with probability at least $1-\alpha$. We present two approaches to constructing prediction sets: (1) CP in Section \ref{subsec:cp}, and (2) likelihood-based prediction in Section \ref{subsec:likelihood}, with more attention afforded to the former.

\subsection{Conformal Prediction}\label{subsec:cp}
As noted before, the methodology described here is a reflection of the work of \cite{nettasinghe2023extending}; however, rather than considering doubly-stochastic processes (like with HMMs), we take a simpler approach making predictions for a single observed time series. Applying the notion of exchangeability to the individual observations of a Markov process is not feasible given the inherent probabilistic dependencies over time. Consequently, we can leverage the idea of \textit{partial-exchangeability} which applies the exchangeability principle, not to each observed time point, but to the entire state-sequence. In particular, given two realized state-sequences of a discrete-time, discrete-state Markov process (of order 1) $\{x_t\}_{t=1}^T$ and $\{x_t'\}_{t=1}^T$, if (1) $x_1 = x_1'$, and (2) the number of observed transitions between states is the same in $\{x_t\}_{t=1}^T$ and $\{x_t'\}_{t=1}^T$, then these sequences must be partially exchangeable in the sense that $\mathbb{P}\qty(\{X_t = x_t\}_{t=1}^T) = \mathbb{P}\qty(\{X_t = x_t'\}_{t=1}^T)$ \citep{nettasinghe2023extending}. 

\begin{algorithm}[!htb]
\scriptsize
\caption{CP algorithm for discrete-state Markov process, adapted from \cite{nettasinghe2023extending}}\label{alg:cpAlg}
\begin{algorithmic}
\spacingset{0.86}
    \scriptsize
    \State \parbox[t]{0.95\linewidth}{%
    \textbf{Input}: Calibration data $\{X_t\}_{t=1}^T$, miscoverage $\alpha \in [0, 1]$, maximum number of permutations $n$ \\}
    
    \State \parbox[t]{0.95\linewidth}{%
    \textbf{Output}: Prediction set $\mathcal{C}_{1-\alpha} \subseteq \mathcal{X}^{T_1}$, where $\mathcal{X}^{T_1}$ is the set of all possible state-sequences of length $T_1$\\}
    
    \For{$\vb*{x}=(x_{T+1},...,x_{T+T_1}) \in \mathcal{X}^{T_1}$} \\
    
    \State 1. Set $X_{T+1}=x_{T+1},\; \hdots,\; X_{T+T_1} = x_{T+T_1}$\\
    
    \State 2. \parbox[t]{0.95\linewidth}{%
        Using $\{X_t\}_{t=1}^{T+T_1}$, estimate the transition probability matrix ($\vb*{P}$) by
    }
    \State 
    \begin{center}
        $\displaystyle
            \begin{aligned}
            \qty[\widehat{\vb*{P}}]_{i,j} &= \frac{\sum_{t=1}^{T+T_1-1} \mathbf{1}(X_t=i \land X_{t+1}=j)}{\sum_{t=1}^{T+T_1-1} \mathbf{1}(X_t=i)}
            \end{aligned}
        $
    \end{center}
    \State \vspace{0.15em}
    
    \State 3. \parbox[t]{0.95\linewidth}{%
      Determine the set of permutable $i$-blocks, $\mathcal{I} = \{I_1, \hdots, I_D\}$, given $X_{T+T_1} = x_{T+T_1} = i$.\\
    }
    
    \State 4. \parbox[t]{0.95\linewidth}{%
      Let $\mathcal{P}$ be the permutation group of $\mathcal{I}$ where each $\rho \in \mathcal{P}$ results in $\{I_1^{(\rho)}, \hdots, I_D^{(\rho)}\}$. Then
      $$\{X_t^{(\rho)}\}_{t=1}^{T+T_1} = \begin{cases}
    \{I_1^{(\rho)}, \hdots, I_D^{(\rho)}, I_{D+1}\}, & X_1 = i\\
    \{I_0, I_1^{(\rho)}, \hdots, I_D^{(\rho)}, I_{D+1}\}, & \text{otherwise}
    \end{cases}$$
    Let $\vb{I}$ denote the original, un-permuted sequence $\{X_t\}_{t=1}^{T+T_1}$ (i.e., ``identity permutation'')
    }
    \State \vspace{0.15em}
    
    \State 5. For computational purposes
    \State \begin{center}
        $\displaystyle
        \Pi :=
        \begin{cases}
            \mathcal{P}, & \text{if } |\mathcal{P}| \leq n \\[0.5em]
            \text{a random sample without replacement of size } n \text{ from } \mathcal{P}, & \text{if } |\mathcal{P}| > n
        \end{cases}$
    \end{center}
    \Statex \vspace{0.15em}
    
    \For{$\pi \in \Pi$}\\
        \State Calculate the non-conformity score $S(\pi)$
        \State \parbox[t]{0.9\linewidth}{%
            $$S(\pi)=1-\frac{\sum_{j=1}^{T_1}\mathbb{P}_{\widehat{P}}\big(X^{(\pi)}_{T+j}\mid X^{(\pi)}_{T}\big)}{T_1}$$
        }        
    \EndFor\\
    
    \State 6. \parbox[t]{0.9\linewidth}{%
        Calculate the p-value 
        $$\widehat{q}(\vb*{x}) = \frac{\sum_{\pi \in \Pi} \mathbf{1}\{S(\pi) > S(\vb{I})\} + u \cdot \mathbf{1}\{S(\pi)=S(\vb{I})\}}{|\Pi|} ,$$
        where $\vb{I}$ is the identity permutation (refer to step 4), and $u \sim \text{Uniform}[0,1]$.\\
    }
    
    \EndFor\\
    
    \State \textbf{return} $\mathcal{C}_{1-\alpha}=\{\vb*{x} \in \mathcal{X}^{T_1}:\widehat{q}(\vb*{x}) > \alpha\}$
\end{algorithmic}
\end{algorithm}

Algorithm \ref{alg:cpAlg} outlines the CP routine for constructing $(1-\alpha)$-level prediction sets for the forecasted state-sequences of length $T_1$. The following is a detailed overview of the mechanics of the algorithm, specifically focusing on how the partial-exchangeability assumption manifests. Based on the work of \cite{chernozhukov2018exact, chernozhukov2021exact}, the idea presented in \cite{nettasinghe2023extending} is to test if a proposed sequence $(x_{T+1}, \hdots, x_{T+T_1})$ belongs in the prediction set based on the exchangeability of permutable segments of the sequence, $\{X_t\}_{t=1}^{T+T_1}$, where it is assumed $X_{T+1} = x_{T+1}, \hdots, X_{T+T_1} = x_{T+T_1}$ (refer to $\{X_t\}_{t=1}^{T+T_1}$ as the \textit{augmented} sequence). In particular, the authors break the augmented sequence into what they term ``$i$-blocks,'' which are sequences of state-observations that begin with $i \in \mathcal{X}$ and include no further $i$'s. For the purposes of Algorithm \ref{alg:cpAlg}, the choice of $i$ is completely determined by the proposed value for $X_{T+T_1}$. With these $i$-blocks, the augmented sequence $\{X_t\}_{t=1}^{T+T_1}$ can be represented as
\begin{equation}\label{eq:iblock}
    \{X_t\}_{t=1}^{T+T_1} = \begin{cases}
    \{I_1, ..., I_D, I_{D+1}\}, & \text{if }\, X_1 = i\\
    \{I_0, I_1, ..., I_D, I_{D+1}\}, & \text{otherwise}
    \end{cases}
\end{equation}
% (i.e., the last time point in the augmented sequence)
where $I_d$, $d \in \{1,2,\hdots, D\}$, denotes a single permutable $i$-block and $D$ is the total number of $i$-blocks. Note that $I_0$ and $I_{D+1}$ are \textit{not} considered $i$-blocks because these cannot be permuted without changing the counts of state transitions of the original, un-permuted sequence $\{X_t\}_{t=1}^{T+T_1}$ (i.e., the \textit{identity permutation}). Define $\mathcal{I} := \{I_1, \hdots, I_D\}$ to be the set of all permutable $i$-blocks. In order to better understand the two cases in (\ref{eq:iblock}), consider the two following simple examples of breaking a state-sequence into $i$-blocks where $\mathcal{X} = \{1,2,3,4\}$, and the only difference is the state at the last time point, namely $X_{T+T_1} = x_{T+T_1} = i$.
\begin{center}
\begin{tikzpicture}[font=\ttfamily, every node/.style={inner sep=1pt},scale=0.8]

%% --- Left Panel --- %%
\begin{scope}
    % Panel Title
    \node at (0,1.0) {\textbf{Example 1} $(i = 1)$};

    % Sequence
    \node (a1) at (-3,0) {1};
    \node (a2) at (-2.175,0) {2};
    \node (a3) at (-1.35,0) {4};
    \node (a4) at (-0.525,0) {1};
    \node (a5) at (0.3,0) {2};
    \node (a6) at (1.125,0) {3};
    \node (a7) at (1.95,0) {4};
    \node (a8) at (2.775,0) {1};

    % Braces for bottom sequence
    \draw [decorate,decoration={brace,amplitude=5pt,mirror}]
        ($(a1.south west)+(0,-0.2)$) -- ($(a3.south east)+(0,-0.2)$)
        node[midway,below=6pt] {$I_1$};

    \draw [decorate,decoration={brace,amplitude=5pt,mirror}]
        ($(a4.south west)+(0,-0.2)$) -- ($(a7.south east)+(0,-0.2)$)
        node[midway,below=6pt] {$I_2$};

    \draw [decorate,decoration={brace,amplitude=5pt,mirror}]
        ($(a8.south west)+(0,-0.2)$) -- ($(a8.south east)+(0,-0.2)$)
        node[midway,below=6pt] {$I_3$};

    % Optional border
    \draw[rounded corners] (-3.8,1.3) rectangle (3.8,-1.2);
\end{scope}

%% --- Right Panel --- %%
\begin{scope}[shift={(10,0)}]
    % Panel Title
    \node at (0,1.0) {\textbf{Example 2} $(i=2)$};

    % Sequence
    \node (a1) at (-3,0) {1};
    \node (a2) at (-2.175,0) {2};
    \node (a3) at (-1.35,0) {4};
    \node (a4) at (-0.525,0) {1};
    \node (a5) at (0.3,0) {2};
    \node (a6) at (1.125,0) {3};
    \node (a7) at (1.95,0) {4};
    \node (a8) at (2.775,0) {2};

    % Braces for bottom sequence
    \draw [decorate,decoration={brace,amplitude=5pt,mirror}]
        ($(a1.south west)+(0,-0.2)$) -- ($(a1.south east)+(0,-0.2)$)
        node[midway,below=6pt] {$I_0$};

    \draw [decorate,decoration={brace,amplitude=5pt,mirror}]
        ($(a2.south west)+(0,-0.2)$) -- ($(a4.south east)+(0,-0.2)$)
        node[midway,below=6pt] {$I_1$};

    \draw [decorate,decoration={brace,amplitude=5pt,mirror}]
        ($(a5.south west)+(0,-0.2)$) -- ($(a7.south east)+(0,-0.2)$)
        node[midway,below=6pt] {$I_2$};

    \draw [decorate,decoration={brace,amplitude=5pt,mirror}]
        ($(a8.south west)+(0,-0.2)$) -- ($(a8.south east)+(0,-0.2)$)
        node[midway,below=6pt] {$I_3$};

    % Optional border
    \draw[rounded corners] (-3.8,1.3) rectangle (3.8,-1.2);
\end{scope}
\end{tikzpicture}
\end{center}
In both cases, $D = 2$ and $\mathcal{I} = \{I_1, I_2\}$. For Example 1, there exists two partially exchangeable sequences, namely $\{I_1, I_2, I_3\}$ (the identity permutation) and $\{I_2, I_1, I_3\}$; for Example 2, the partially exchangeable sequences are $\{I_0, I_1, I_2, I_3\}$ (the identity permutation) and $\{I_0, I_2, I_1, I_3\}$. In each example, no matter how we permute $I_1$ and $I_2$, the resulting state-sequence will have the same number of state transitions as the initial, un-permuted augmented sequence. Provided this intuition for partial-exchangeability, we now have the architecture to obtain valid, $(1-\alpha)$-level prediction sets for state-sequences $T_1$ time points into the future (see \cite{nettasinghe2023extending} for proof of validity).

While this partial-exchangeability CP framework serves as the foundation for our paper, alternative approaches exist for applying CP methods to Markovian or temporally dependent data \citep[e.g.,][]{cherubin2016, stankeviciute2021conformal, LeeSTCP, zheng2024conformal, dai2026distributionalconformalpredictionmarkov}. However, these alterative approaches are not directly applicable to our setting in which only a single discrete-state Markov trajectory is observed and the objective is to construct prediction sets for future state-sequences of arbitrary length with finite-sample coverage guarantees. For example, \cite{cherubin2016} and \cite{stankeviciute2021conformal} provide exact coverage guarantees but rely on exchangeability across \textit{multiple} observed time series rather than a single one. Meanwhile, \cite{dai2026distributionalconformalpredictionmarkov} considers single Markov trajectories; however, it focuses only on single-step predictions for continuous-valued processes, as opposed to discrete-valued predictions of arbitrary lengths. That said, we can benchmark the performance of utilizing this partial-exchangeability CP framework by comparing its results to that of more standard CP methods in a simulation study where the data generating mechanism is defined by a discrete-time, discrete-state Markov process. The two alternative CP approaches tested are \textit{classical} CP and \textit{weighted} CP \citep{angelopoulos2021gentle}. The details of these algorithms can be found in Appendix \ref{app:addResults}; to summarize, however, the results provide further evidence supporting Algorithm \ref{alg:cpAlg} and the use of the partial-exchangeability assumption in the presence of temporally dependent data.

\subsection{Likelihood-Based Prediction}\label{subsec:likelihood}
An alternative approach to CP is to simply fit a Markov model to the observed state-sequences in the calibration data and obtain likelihood-based prediction sets for forecasted state-sequences of length $T_1$.  This approach serves as a baseline comparison to the prediction sets obtained using CP. However, unlike CP, this method does not offer finite-sample guarantees that are independent of model correctness (i.e., not robust to model-misspecification).

The likelihood-based prediction sets for individual state-sequences are constructed as follows. We first fit a Markov model by estimating the transition probability matrix $\widehat{\vb*{P}}$ following the estimator from step 2 of Algorithm \ref{alg:cpAlg}, but only with the calibration data, $\{X_t\}_{t=1}^T$. Then, if $\mathcal{X}^{T_1} = \{\vb*{x}^{1}, \hdots, \vb*{x}^{J}\}$ denotes the set of all possible state-sequences of length $T_1$ ($J = |\mathcal{X}^{T_1}|$), we find 
\begin{equation}\label{eq:likeApproach}
\begin{split}
\mathbb{P}(\vb*{X}_{T+1}^{T+T_1} = \vb*{x}^{j} \mid X_T = x_T) &:= \mathbb{P}(X_{T+1} = x^j_{T+1}, \hdots, X_{T+T_1} = x^j_{T+T_1} \mid X_T = x_{T})\\
&= \qty[\widehat{\vb*{P}}]_{x_{T},\, x^j_{T+1}} \cdot \prod_{t = 2}^{T_1} \qty[\widehat{\vb*{P}}]_{x^j_{T+t-1},\, x^j_{T+t}}
\end{split}
\end{equation}
for all $j \in \{1,2,\hdots, J\}$, where $[\cdot]_{k,l}$ denotes the $k^{th}$ row and $l^{th}$ column of a matrix argument. Next, let $\vb*{x}^{(1)}, \vb*{x}^{(2)}, \hdots \vb*{x}^{(J)}\in \mathcal{X}^{T_1}$ denote the ordered elements (greatest to least) of $\mathcal{X}^{T_1}$ based on the probability mass computed in (\ref{eq:likeApproach}). The likelihood-based prediction set at level $(1-\alpha)$ is taken to be the highest probability mass set:
\begin{equation}\label{eq:likeDef}
    \mathcal{C}^{\text{like}}_{1-\alpha} := \{\vb*{x}^{(1)}, \hdots, \vb*{x}^{(K)}\},
\end{equation}
where
$$K = \min\qty{k \in \{1,2,\hdots, J\} \; \Bigg| \; \sum_{j = 1}^k \mathbb{P}\qty(\vb*{X}_{T+1}^{T+T_1} = \vb*{x}^{(j)} \mid X_T=x_T) \geq 1-\alpha}.$$
In the case of ties where, say, $\mathbb{P}(\vb*{X}_{T+1}^{T+T_1} = \vb*{x}^{\ell} \mid X_T=x_T) = \mathbb{P}(\vb*{X}_{T+1}^{T+T_1} = \vb*{x}^{j} \mid X_T=x_T)$ for $\ell, j \in \{1,\hdots, J\}$, $\ell \neq j$, the ordering of $\vb*{x}^{\ell}$ and $\vb*{x}^{j}$ is determined uniformly at random. This handling of ties is analogous to sampling $u\sim \text{Uniform}[0,1]$ in step 6 of Algorithm \ref{alg:cpAlg} for CP. The reason we use only the calibration data to estimate $\widehat{\vb*{P}}$ as opposed to the augmented sequence, $\{X_t\}_{t=1}^{T+T_1}$, is because the augmented sequence would lead to a different fitted model for each proposed future sequence. In other words, we would have $\widehat{\vb*{P}}_j$ for each $\vb*{x}^{j}\in \mathcal{X}^{T_1}$, thus changing (\ref{eq:likeApproach}) to $\mathbb{P}(\vb*{X}_{T+1}^{T+T_1} = \vb*{x}^{j} \mid X_T = x_T, \; \widehat{\vb*{P}}_j)$. Therefore, the resulting candidate-specific probabilities in (\ref{eq:likeApproach}) would, in general, not correspond to a probability mass under a \textit{common} fitted probability measure, making the interpretation of $\mathcal{C}^{\text{like}}_{1-\alpha}$ in (\ref{eq:likeDef}) ill-posed because we cannot guarantee $\sum_{j = 1}^J \mathbb{P}\qty(\vb*{X}_{T+1}^{T+T_1} = \vb*{x}^{(j)} \mid X_T=x_T, \; \widehat{\vb*{P}}_{(j)}) = 1$.

Next, the motivation for constructing the prediction set $\mathcal{C}^{\text{like}}_{1-\alpha}$ in (\ref{eq:likeDef}) follows from the ``highest-density region'' literature \citep{hyndman1995highest, hyndman1996computing}. Albeit these foundational references are applicable to continuous responses, the same procedure can be extended for discrete data \citep{o2022smallest}, as well as temporally dependent data \citep{mukhopadhyay2019predictive, homburg2021performance}. In this discrete, finite setting, the construction of (\ref{eq:likeDef}) shares many characteristics to that of the classic ``0-1 knapsack problem,'' initially proposed in \cite{mathews1896partition}. While (\ref{eq:likeDef}) serves as a baseline comparison to CP, it is worth noting that alternative approaches exist for constructing likelihood-based prediction sets (see Section \ref{app:like}). Our motivation for choosing (\ref{eq:likeDef}), however, is that it guarantees the minimal cardinality over all prediction sets having cumulative probability mass of at least $1-\alpha$.

\subsubsection{Alternative Likelihood-Based Prediction Strategies} \label{app:like}
The likelihood-based prediction set defined in (\ref{eq:likeDef}) minimizes the cardinality of the $(1-\alpha)$-level prediction sets, and guarantees that \textit{each} $\mathcal{C}^{\text{like}}_{1-\alpha}$ obtains a cumulative probability mass of at least $1-\alpha$. However, as we will show in Figures \ref{fig:rel_curves_mc} and \ref{fig:rel_curves_fatal}, the empirical coverage from this likelihood-based approach tends to exceed, on average, the desired coverage level. As a result, we present alternative likelihood-based prediction strategies that result in better calibrated coverage, on average, at the sacrifice of \textit{not} guaranteeing each prediction set obtains cumulative probability mass of at least $1-\alpha$.

Let $\mathcal{C}^{\text{like}^*}_{1-\alpha}$ denote one alternative likelihood-based prediction set for desired coverage level $1-\alpha$. We follow all of the steps in Section \ref{subsec:likelihood} leading up to the construction of the prediction set; however, now the prediction set is defined as
\begin{equation}\label{eq:cLikeStar}
    \mathcal{C}^{\text{like}^*}_{1-\alpha} := \{\vb*{x}^{(1)}, \hdots, \vb*{x}^{(K)}\}\cdot \mathbf{1}\{u^* \leq p\} + \{\vb*{x}^{(1)}, \hdots, \vb*{x}^{(K-1)}\}\cdot \mathbf{1}\{u^* > p\},
\end{equation}
where $u^*$ is a realization of $U^*\sim \text{Uniform}[0,1]$, and
{\scriptsize
$$p = \begin{cases}
    \frac{1\; - \;\alpha}{\mathbb{P}\qty(\vb*{X}_{T+1}^{T+T_1} \; = \; \vb*{x}^{(K)}\; \mid \; X_T\; = \;x_T)}, & K=1\\
    \frac{(1\; - \;\alpha) \; - \; \sum_{j = 1}^{K-1} \mathbb{P}\qty(\vb*{X}_{T+1}^{T+T_1} \; = \; \vb*{x}^{(j)} \; \mid \; X_T\; = \;x_T)}{\mathbb{P}\qty(\vb*{X}_{T+1}^{T+T_1} \; = \; \vb*{x}^{(K)} \; \mid \; X_T\; = \;x_T)}, & K > 1.
\end{cases}$$
}%
This choice of $p$ guarantees that the {\em expected} cumulative probability mass from $\mathcal{C}^{\text{like}^*}_{1-\alpha}$ will equal $1-\alpha$ (proof in Appendix \ref{app:likeProof}).

The reason we ultimately elect \textit{not} to implement this is because, again, it no longer guarantees that $\mathcal{C}^{\text{like}^*}_{1-\alpha}$ is a $1-\alpha$ probability set with respect to the probability distribution over the space of sequences.  Of course, on average it will be calibrated to $1-\alpha$ coverage, but its interpretation may appear suspicious to practitioners.  To see this, consider an extreme alternative where $\mathcal{C}^{\text{like}^{**}}_{1-\alpha} :=\mathcal{X}^{T_1} \cdot \mathbf{1}\{u^* \leq 1-\alpha\} + \emptyset \cdot \mathbf{1}\{u^* > 1-\alpha\}$. This likelihood-based prediction set is perfectly calibrated to $1-\alpha$ \textit{on average}; however, the set is either all possible forecasted state-sequences ($\mathcal{X}^{T_1}$ with $\mathbb{P}(\mathcal{X}^{T_1}) = 1$) or the empty-set ($\emptyset$ with $\mathbb{P}(\emptyset) = 0$).

\section{Experiments and Evaluation}\label{sec:experiments}

\subsection{Data Description and Cleaning}\label{sec:data_clean}
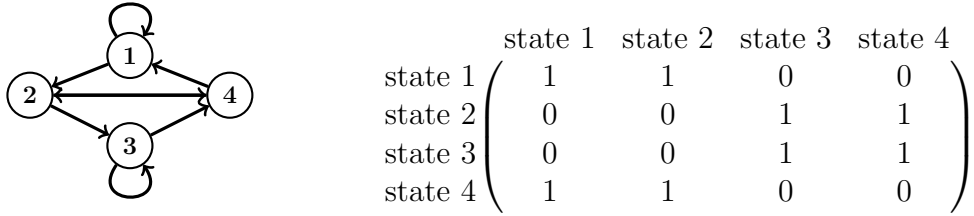
\begin{figure}[!htb]
\centering

\begin{minipage}{0.35\textwidth}
    \centering
    \begin{tikzpicture}[scale=0.9, transform shape, node distance={15mm}, thick, main/.style = {draw, circle}] 
    \node[main] (1) {\textbf{1}}; 
    \node[main] (2) [below left of=1, yshift = 4mm, xshift=-6mm]{\textbf{2}}; 
    \node[main] (3) [below of=1] {\textbf{3}}; 
    \node[main] (4) [below right of=1, yshift = 4mm, xshift=6mm] {\textbf{4}}; 
    \draw[very thick, ->] (1) -- (2); 
    \draw[very thick, ->] (2) -- (3);
    \draw[very thick, ->] (3) -- (4);
    \draw[very thick, ->] (2) -- (4); 
    \draw[very thick, ->] (4) -- (2); 
    \draw[very thick, ->] (4) -- (1); 
    \draw[very thick, ->] (1) to [out=125,in=55,looseness=5] (1); 
    \draw[very thick, ->] (3) to [out=235,in=305,looseness=5] (3);   
    \end{tikzpicture}
\end{minipage}
\hspace{2em}
\begin{minipage}{0.35\textwidth}
    \centering
    $$\begin{pNiceMatrix}[first-row,first-col]
    & \text{state 1} & \text{state 2} & \text{state 3} & \text{state 4}\\
\text{state 1}   & 1    & 1    & 0    & 0 \\
\text{state 2}   & 0    & 0    & 1    & 1 \\
\text{state 3}   & 0    & 0    & 1    & 1 \\
\text{state 4}   & 1    & 1    & 0    & 0 
\end{pNiceMatrix}$$
\end{minipage}

\caption{\footnotesize All allowable state transitions between the four conflict-states (left), and the corresponding adjacency matrix (right), where states 1, 2, 3, and 4 correspond to peacetime, escalation, war, and deescalation, respectively.}
\label{fig:stateTransitions}
\end{figure}

The data for this work comes from Uppsala Conflict Data Programme (UCDP), which tracks fatalities from organized violence across the world \citep{davies2025organized}. In particular, we utilize the UCDP Georeferenced Event Dataset (GED) Global, and aggregate the data to the country-month level \citep{sundberg2013introducing}. Each observation corresponds to the number of fatalities from state-based conflict observed for a given country in a given month. In total, we utilize data from 191 countries from January of 1990 until July of 2025. Building on the Markov structure proposed by \citet{randahl2022predicting, randahl2024forecasting}, we deterministically assign state-labels to these counts in order to categorize the type of conflict-state that a given country is in at any given month. We will consider four states: (1) peacetime, (2) escalation, (3) war, and (4) deescalation. Let $Y_{i, t}$ denote the number of casualties observed for country $i$, $i \in$ $\{1,2\hdots, 191\}$, at month $t$. The assigned state-label, $X_{i,t}$, is determined by the following rules: 
$$
X_{i,t} = \begin{cases}
    1, & \text{if } \; Y_{i,t-1} = Y_{i,t} = 0 \; \text{ for } \; t > 1, \; \text{ or } \; Y_{i,t} = 0 \; \text{ for } \; t = 1\\
    2, & \text{if } \; Y_{i,t-1} = 0 \; \text{ and } Y_{i,t} > 0 \; \text{ for } \; t > 1\\
    3, & \text{if } \; Y_{i,t-1} > 0 \; \text{ and } Y_{i,t} > 0 \; \text{ for } \; t > 1, \; \text{ or } \; Y_{i,t} > 0 \; \text{ for } \; t = 1\\
    4, & \text{if } \; Y_{i,t-1} > 0 \; \text{ and } Y_{i,t} = 0 \; \text{ for } \; t > 1.
\end{cases}
$$
Note that the only data used in the analysis are these assigned state-labels over time. From these state-labels, Figure \ref{fig:stateTransitions} illustrates the allowable state transitions and corresponding adjacency matrix.

Next, we clean the data according to the following rule: the countries kept in the analysis must have at least \textit{five} observations that are not in peacetime (i.e., at least five observations of states 2, 3, and/or 4), and the proportion of instances of peacetime (state 1) does not exceed 0.99 for a given country. In summary, the goal of this rule is to exclude countries that have the same state observed across all (or almost all) time points, and in these data, we find this occurs often with countries in peacetime (state 1). The only exception to this is Ukraine, which remains in war (state 3) for all observations, thus we also exclude it from our analysis. Additional insight which motivates this data-cleaning procedure is found in Section \ref{sec:limitations}. After cleaning the data, we are left with 86 countries in our analysis.

\subsection{Simulation Study}\label{sec:sim}
In order to evaluate the methods discussed in Section \ref{sec:method}, we present a Monte Carlo simulation study where the true data generating mechanism is known. The data are simulated as a Markov process where the true transition probability matrix, $\vb*{P}$, and initial state distribution, $\vb*{\pi}$, are given as
\begin{equation}\label{eq:simGen}
    \vb*{P} =
    \mqty(
    0.895 & 0.105 & 0 & 0 \\
    0 & 0 & 0.500 & 0.500 \\
    0 & 0 & 0.722 & 0.278 \\
    0.653 & 0.347 & 0 & 0
    ),
    \quad
    \vb*{\pi} =
    \mqty(
    0.25 & 0.25 & 0.25 & 0.25
    ).
\end{equation}
These true values are informed by the cleaned, real data. In particular, using the estimator from step 2 of Algorithm \ref{alg:cpAlg}, we compute an estimated transition probability matrix for each country, and then average (component-wise) the estimated transition probability matrices to determine the values in $\vb*{P}$. For the true initial state probabilities, $\vb*{\pi}$, we assign equal probability to starting in any one of the four states. 

We can evaluate the CP and likelihood-based approaches by comparing both their reliability curves as well as the composition of their prediction sets for prespecified confidence levels and forecast lengths. For each simulation, $T = 200$ is the length of the calibration sequence. Next, let $T_1 \in \{1,2,\hdots, 6\}$ denote the test length of the forecasted state-sequence. Then, in order to compare the empirical coverages of the CP and likelihood-based approaches for each $T_1$, we run the simulation for a range of confidence levels, namely for $1-\alpha \in \{0.50, 0.55, \hdots, 1.00\}$. For each combination of $T_1$ and $1-\alpha$, we simulate 500 state-sequences of length $T + T_1$ according to the Markov process defined by (\ref{eq:simGen}). The CP and likelihood-based prediction sets are constructed using the first $T$ time points (the calibration data), and then for each of the 500 simulations, we can determine whether or not the prediction sets [for the last $T_1$ time points] contain the true forecasted state-sequence. The empirical coverage is then determined as the proportion of the 500 prediction sets that contain the true forecasted state-sequence.
\begin{figure}[!htb]
    \centering
    \includegraphics[width=0.35\linewidth, trim={0cm 0.5cm 0cm 0.9cm},clip]{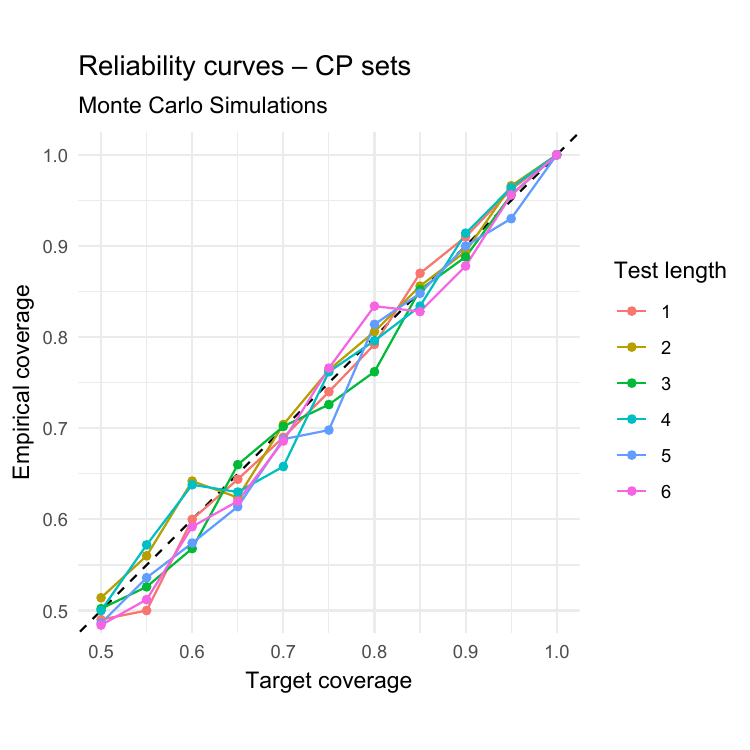}
    \includegraphics[width=0.35\linewidth, trim={0cm 0.5cm 0cm 0.9cm},clip]{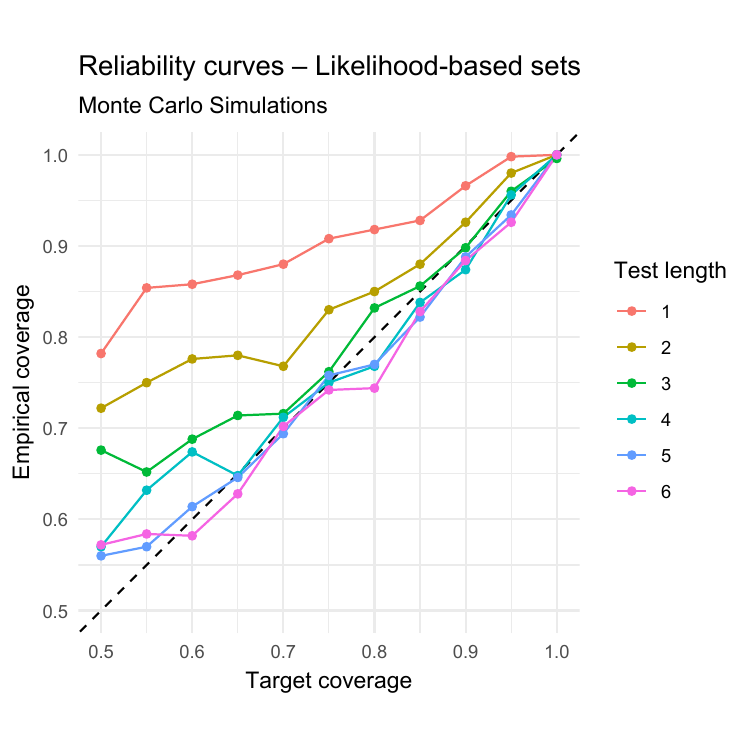}
    \caption{\footnotesize Reliability curves illustrating the relationship between the empirical coverage and the target coverage for both the CP approach (left) and likelihood-based prediction approach (right) for the simulated data.}
    \label{fig:rel_curves_mc}
\end{figure}

Figure \ref{fig:rel_curves_mc} presents the reliability curves for the conformal and likelihood-based predictions. We see that the CP approach remains well calibrated for all confidence levels and test lengths, consistent with the results in \cite{nettasinghe2023extending}. In the case of the likelihood-based approach, however, we see that as both the test length and confidence level decrease, the prediction sets tend to \textit{over-cover}. Insight into the over-coverage exhibited by the likelihood-based approach is provided in Section \ref{subsec:overcovg}.

\begin{figure}[!htb]
    \centering
    \includegraphics[width=0.7\linewidth, trim={0cm 0.5cm 0cm 0.5cm},clip]{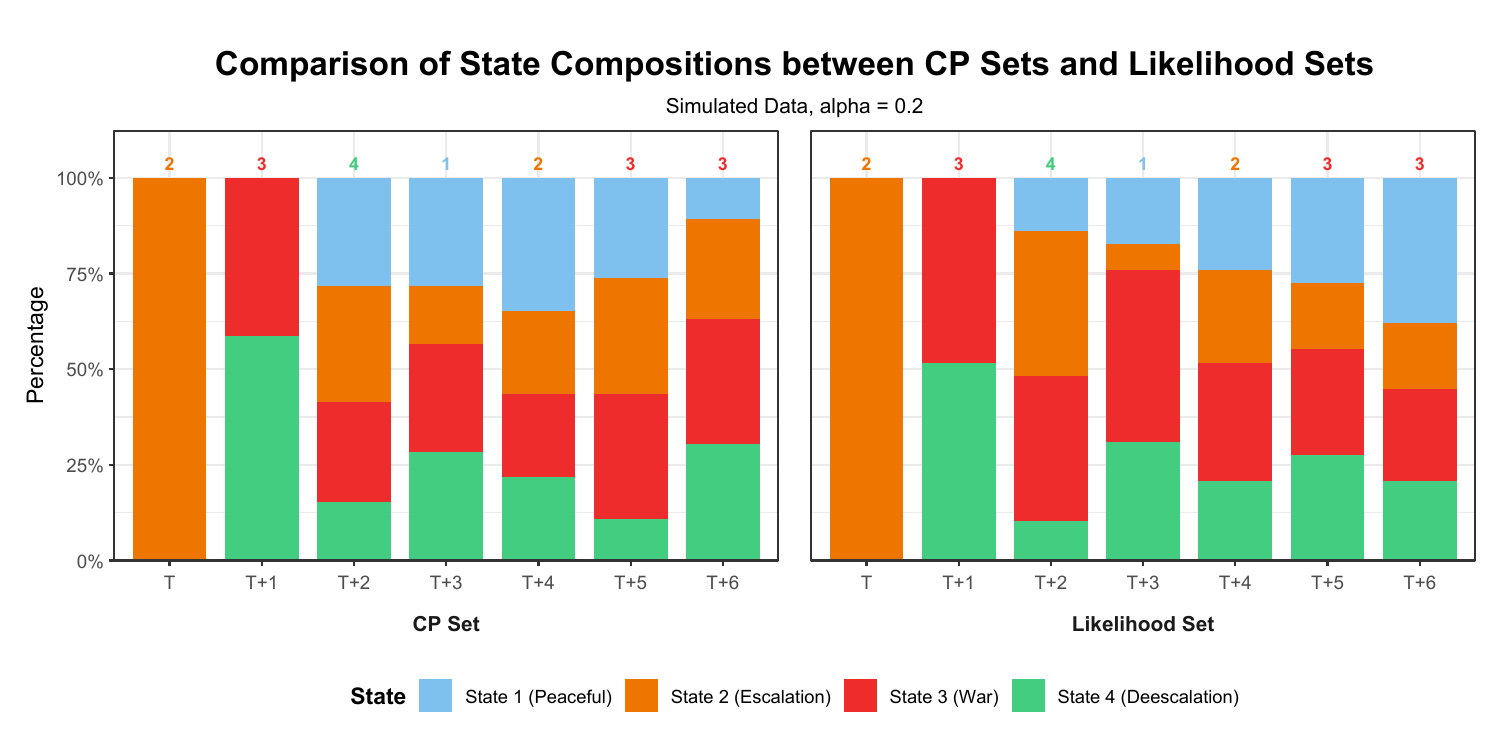}
    \caption{\footnotesize Visualization of the prediction sets for a single simulated state-sequence resulting from the conformal (left) and likelihood-based (right) prediction approaches. Note that the state at time $T$ (i.e., $X_T$) is \textit{not} being forecasted; however, the realized value of $X_T$ is being plotted because the possible forecasted state-sequences ($X_{T+1}, \hdots, X_{T+T_1}$) depend on $X_T$ by way of the Markov assumption.}
    \label{fig:set_vis_mc}
\end{figure}
Next, Figure \ref{fig:set_vis_mc} is a visualization of the prediction sets for a single simulated state-sequence where $T=200$, $T_1 = 6$, and $1-\alpha = 0.80$, using both the conformal and likelihood-based methods. Figure \ref{fig:set_vis_mc} has the following interpretation: taking all state-sequences in the prediction set, the proportion of each state at each time point ($T+1, T+2, \hdots, T+T_1$) is represented by the proportion of each corresponding color in each vertical bar. In this case, the true forecasted state for each future time point is denoted by the colored numeral above each vertical bar. The results in Figure \ref{fig:set_vis_mc} illustrate an important difference between the CP and likelihood-based approaches. We see that as the test length, $T_1$, increases, the composition of the prediction sets produced from the likelihood-based approach converges to its stationary distribution; however, the CP approach produces a more diverse set of forecasted state-sequences, with a difference more clearly pronounced for the later time points (e.g., $T+6$). For example, given $X_T = 2$ in Figure \ref{fig:set_vis_mc} and given the true model in (\ref{eq:simGen}), we can estimate the stationary distribution $P(X_{T+\infty} \mid X_T = 2) = (0.6206, 0.0999, 0.1796, 0.0999)$; this conditional probability mass is close to the composition of predicted states at time $T+6$ for the likelihood-based approach.

Lastly, Table \ref{tab:cardsets} presents the differences in the cardinality of the prediction sets between the conformal and likelihood-based prediction algorithms. On average, the likelihood-based prediction sets contain fewer predicted state-sequences than the CP sets (with exception to $T_1=1$ and target coverage $0.55$). 
As mentioned in Section \ref{subsec:likelihood}, the likelihood-based prediction sets are constructed to minimize the cardinality of the prediction set while guaranteeing a cumulative probability mass of at least $1-\alpha$. 
% As mentioned in Section \ref{subsec:likelihood}, the likelihood-based prediction sets are constructed as efficiently as possible given that the model is correctly specified in the simulation study, in the sense of minimizing the number of forecasted state-sequences while guaranteeing a cumulative probability mass of at least $1-\alpha$. 
Although the greater cardinality suggests that CP is less efficient than the likelihood-based approach, Figure \ref{fig:rel_curves_mc} indicates that the likelihood-based approach tends to over-cover due to fewer sequences accounting for most of the probability mass.
\begin{table}[t]
\centering
\small
\begin{tabular}{c*{5}{rr}}
& \multicolumn{10}{c}{\textbf{Target Coverage}} \\
\cmidrule(lr){2-11}
& \multicolumn{2}{c}{0.55}
& \multicolumn{2}{c}{0.65}
& \multicolumn{2}{c}{0.75}
& \multicolumn{2}{c}{0.85}
& \multicolumn{2}{c}{0.95} \\
\cmidrule(lr){2-11}
\phantom{Horizon}
& CP & Like.
& CP & Like.
& CP & Like.
& CP & Like.
& CP & Like. \\
\midrule

$(T_1 = 1)$ & 0.86 & 1.04 & 1.23 & 1.19 & 1.51 & 1.36 & 1.61 & 1.36 & 1.87 & 2.00 \\
$(T_1 = 2)$ & 1.47 & 1.18 & 1.84 & 1.50 & 2.26 & 1.81 & 2.67 & 2.32 & 3.63 & 3.45 \\
$(T_1 = 3)$ & 2.68 & 1.41 & 3.06 & 1.69 & 3.97 & 2.74 & 4.81 & 3.91 & 6.76 & 6.15 \\
$(T_1 = 4)$ & 3.76 & 2.15 & 5.09 & 2.61 & 6.74 & 4.10 & 9.58 & 6.46 & 13.46 & 10.75 \\
$(T_1 = 5)$ & 6.12 & 2.57 & 9.07 & 4.05 & 13.44 & 6.82 & 18.37 & 10.34 & 25.79 & 18.78 \\
$(T_1 = 6)$ & 11.28 & 4.14 & 17.16 & 7.16 & 24.50 & 10.35 & 35.51 & 18.23 & 50.19 & 33.87 
\end{tabular}
\caption{\footnotesize Comparison of the cardinality of prediction sets using CP and likelihood-based prediction (``Like.'') for the simulation over different prediction lengths, $T_1$, and target coverage levels. The results are averaged over 100 simulations.}
\label{tab:cardsets}
\end{table}

\subsubsection{Insight into Over-Coverage for Likelihood-Based Approach} \label{subsec:overcovg}
In order to gain intuition for the over-coverage observed in the likelihood-based approach in Figure \ref{fig:rel_curves_mc}, let us focus on the simplest case where $T_1 = 1$ and the target coverage is $1-\alpha = 0.50$. Recall how the likelihood-based prediction set, $\mathcal{C}^{\text{like}}_{1-\alpha}$, is defined from Section \ref{subsec:likelihood}. Given the state at time $T$ ($X_T = x_T$) and $T_1 = 1$, we know that $\mathcal{C}^{\text{like}}_{1-\alpha}$ will comprise of either only one or two states. This certainty comes from the fact that in each row of the adjacency matrix in Figure \ref{fig:stateTransitions}, there only ever exists two non-zero entries. From the true model in (\ref{eq:simGen}), the maximum transition probability in each row of $\vb*{P}$ is at least $0.50$, and recall we are considering the case when the target coverage is $1- \alpha = 0.50$. Note that while the true $\vb*{P}$ is never used in the construction of $\mathcal{C}^{\text{like}}_{1-\alpha}$, we know $\widehat{\vb*{P}} \approx \vb*{P}$ on average because $T =200$ is sufficiently large. Given this setup, we know that for a target coverage $1-\alpha = 0.50$, the average cardinality of $\mathcal{C}^{\text{like}}_{1-\alpha}$ is one. This is because the forecasted state-sequence of length $T_1 = 1$ in $\mathcal{C}^{\text{like}}_{1-\alpha}$ is simply the state corresponding to the maximum transition probability in the $x_T$ row of $\widehat{\vb*{P}}$ which, on average, is at least $1-\alpha = 0.50$.

While the empirical coverage is computed according to the method outlined in Section \ref{sec:sim}, we can estimate the coverage for this simple case of $T_1 = 1$ and $1-\alpha =0.50$ using the true model parameter values in ($\ref{eq:simGen}$). From the above discussion, the estimated empirical coverage probability for any $X_T = x_T$ is given by the maximum value in each row of $\vb*{P}$, namely 
$\vb*{p}_{max} := \mqty(0.895, & 0.500, & 0.722, & 0.653)^T.$
Then, in order to obtain a single estimated empirical coverage probability, we need to average the components of $\vb*{p}_{max}$ proportional to the marginal probability distribution of $X_T$, for $T=200$. Using the true model (\ref{eq:simGen}):
\begin{center}
    \begin{tabular}{c|c c c c} 
    & $x_T = 1$ & $x_T = 2$ & $x_T = 3$ & $x_T = 4$\\
    \hline
    $P(X_T = x_t)$ & 0.6206 & 0.0999 & 0.1796 & 0.0999
\end{tabular}
\end{center}
Define $\vb*{p}_{marg} := \mqty(0.6206, & 0.0999, & 0.1796, & 0.0999)^T$. Therefore, we would expect the empirical coverage to be approximately $(\vb*{p}_{max})^{T} \vb*{p}_{marg} = 0.800$. Then, in the right panel of Figure \ref{fig:rel_curves_mc}, if we find the point associated with $T_1 = 1$ and $1-\alpha = 0.50$, we see that it has an empirical coverage of approximately 0.800, thus confirming our reasoning for the over-coverage in the likelihood-based approach here. In essence, the over-coverage can be explained by the discrete-nature of this problem with a small state-space.

\subsection{Real Data Analysis}\label{sec:realDataAnalysis}
After comparing the CP and likelihood-based prediction approaches on synthetic data in Section \ref{sec:sim}, we apply both prediction algorithms to the real conflict data. We can split this real data analysis into two parts. First, for each country, we use a subset of the observations as calibration data, and then test if the true forecasted state-sequence (defined as the next $T_1$ time points following the last time point of the calibration data) is within either methods' prediction sets. We then gauge the efficacy of the two prediction algorithms by comparing the empirical coverage averaged across all countries with a prespecified nominal level (i.e., compare reliability curves similar to Figure \ref{fig:rel_curves_mc}). The second part focuses on constructing prediction sets for a real forecasted state-sequence $T_1$ time points into the future, beyond the time-frame of the observations in the data. In this case, the calibration data for each country is comprised of \textit{all} observations. While we cannot construct reliability curves for this part, we can get a visual understanding of the forecasted prediction sets using visualizations similar to Figure \ref{fig:set_vis_mc}.

\begin{figure}[!htb]
    \centering
    \includegraphics[width=0.35\linewidth, trim={0cm 0.5cm 0cm 0.75cm},clip]{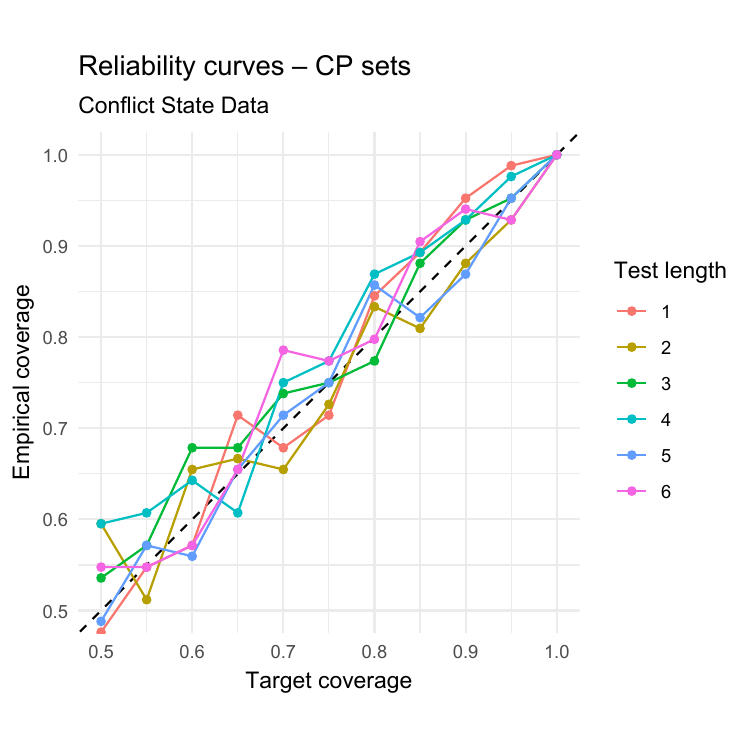}
    \includegraphics[width=0.35\linewidth, trim={0cm 0.5cm 0cm 0.75cm},clip]{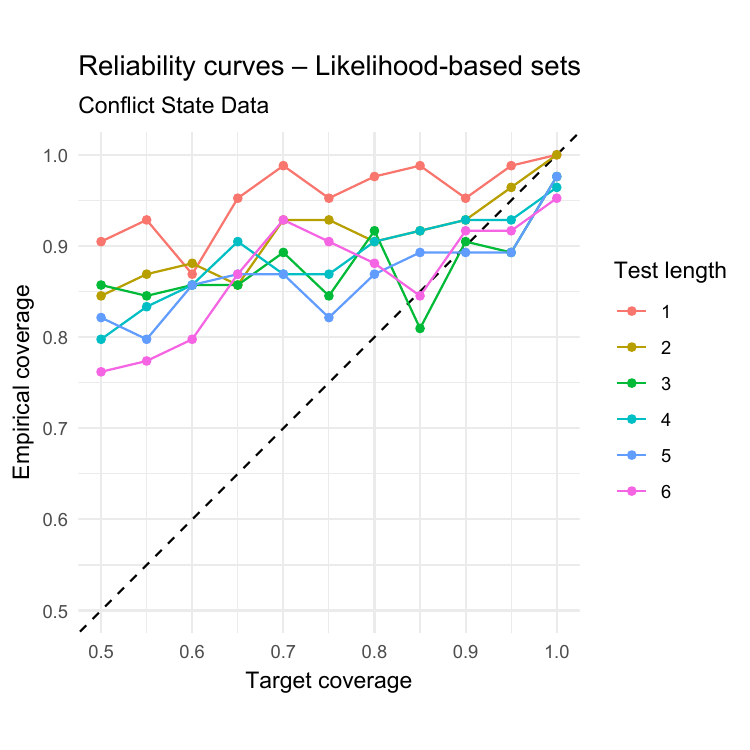}
    \caption{\footnotesize Reliability curves illustrating the relationship between the empirical coverage and the target coverage for both the CP approach (left) and likelihood-based prediction approach (right) for the real conflict data.}
    \label{fig:rel_curves_fatal}
\end{figure}

For the first part of this analysis, Figure \ref{fig:rel_curves_fatal} presents the reliability curves for the two prediction methods applied to the real data where the length of the calibration data, $T$, is assumed to be all data until December 2024. Similar to Section \ref{sec:sim}, we vary the target coverage $1-\alpha \in \{0.50, 0.55, \hdots, 1.00\}$ and test length $T_1 \in \{1,2,\hdots, 6\}$. For each combination of $1-\alpha$ and $T_1$, the empirical coverage is computed as follows: for each country, construct the prediction set for the last $T_1$ time points, determine if that country's true state-sequence is contained within the prediction set, then take the proportion of countries for which the prediction set contains the true sequence. 
From Figure \ref{fig:rel_curves_fatal}, we see that the application of CP leads to an empirical coverage quite close to the nominal coverage for all $T_1$. Contrarily, the likelihood-based approach tends to over-cover for all values of $T_1$ at lower confidence levels, similar to what is seen in Section \ref{sec:sim}. Additionally, if we focus on the special case of $1-\alpha = 1.00$, we see that for the CP approach, the empirical and target coverages are both equal to one for all $T_1$; however, for the likelihood-based approach, there are values of $T_1$ such that the empirical coverage is less than one. The justification for this is as follows. In rare cases, the estimated model, $\widehat{\vb*{P}}$, will assign zero probability to a state-transition that is possible because that specific state-transition is not observed in the calibration data; henceforth, no state-sequences in the likelihood-based prediction set will contain that specific transition, and so if the true forecasted state-sequence contains that state-transition, it will not be contained in the set, resulting in under-coverage for $1-\alpha = 1.00$. Alternatively, this phenomenon is not observed using CP because $\widehat{\vb*{P}}$ (from step 2 of Algorithm \ref{alg:cpAlg}) is computed using the \textit{augmented} state-sequence, not the calibration sequence. This is an important distinction between the CP and likelihood-based predictions, especially if we are interested in forecasting ``abnormal'' events. The likelihood-based approach can \textit{only} forecast state-sequences consistent with state-transitions observed in the calibration data, whereas the CP approach offers more flexibility and can forecast state-sequences that contain state-transitions unobserved in the calibration sequence.

\begin{figure}[!htb]
    \centering
    \includegraphics[width=0.7\linewidth, trim={0cm 0.5cm 0cm 0.5cm},clip]{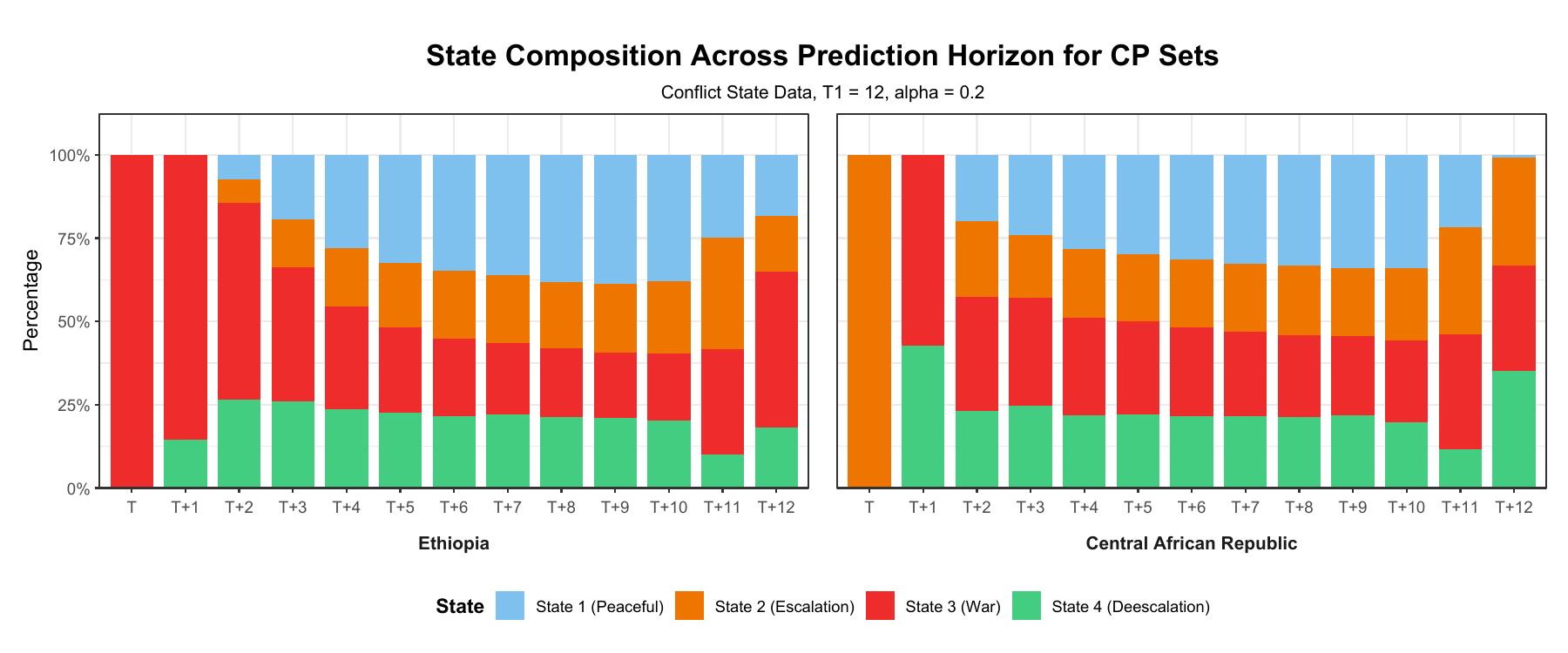}
    \includegraphics[width=0.7\linewidth, trim={0cm 0.5cm 0cm 0.5cm},clip]{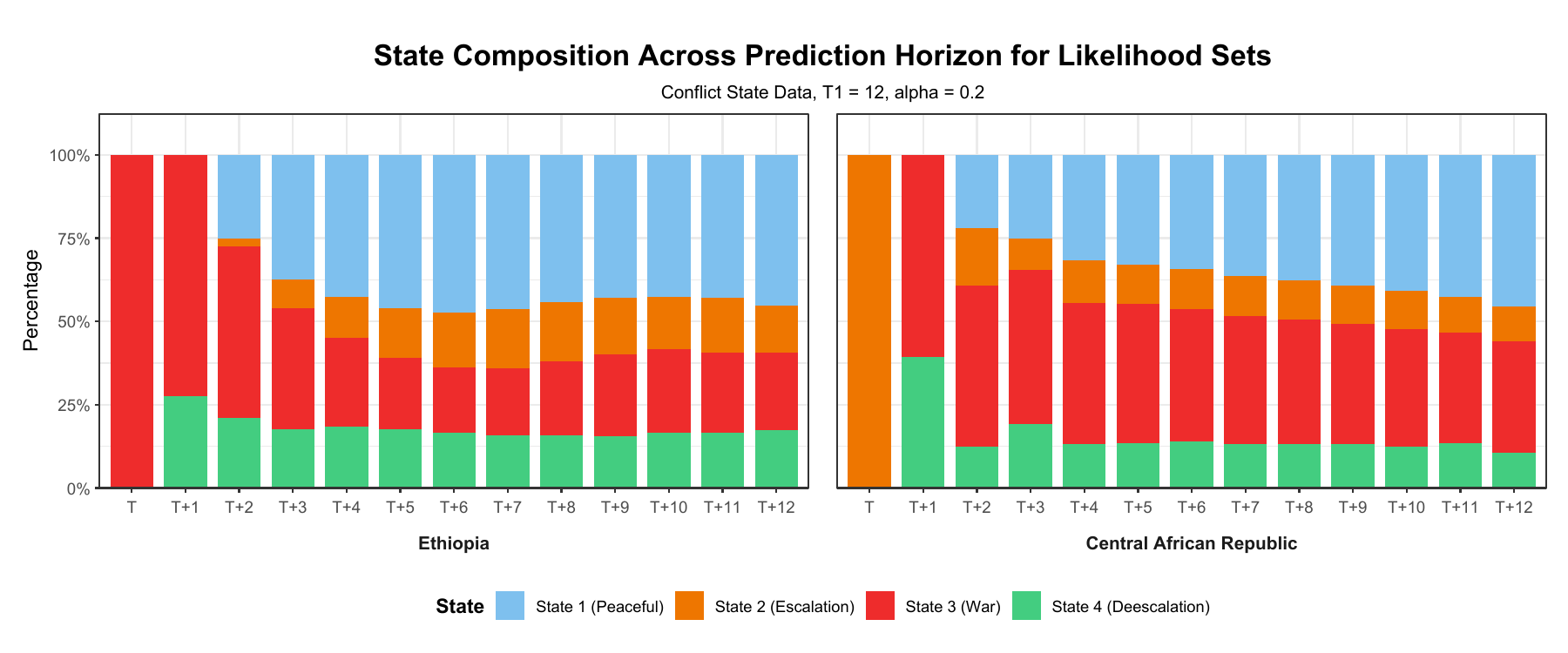}
    \caption{\footnotesize Composition of the prediction sets of forecasted state-sequences, where $T_1 = 12$ and $1-\alpha = 0.80$.}
    \label{fig:predSet12}
\end{figure}

For the second part of this analysis, we use all of the observed data as the calibration data and construct prediction sets for six months beyond the time horizon of our data ($T_1 = 6$) with a confidence level of $1-\alpha = 0.80$. As an empirical illustration of the method and to highlight the differences with the likelihood-based approach, we focus on four specific countries: Central African Republic, Myanmar, Ethiopia, and Colombia. These countries are selected because they exhibit global diversity, as well as diverse and enduring conflict dynamics, such as prolonged, aborted, and intermittent civil wars that stress-test the forecasting approach across meaningfully different Markovian trajectories. 
\begin{figure}[!htb]
    \centering
    \includegraphics[width=0.7\textwidth, trim={0cm 0.5cm 0cm 0.5cm},clip]{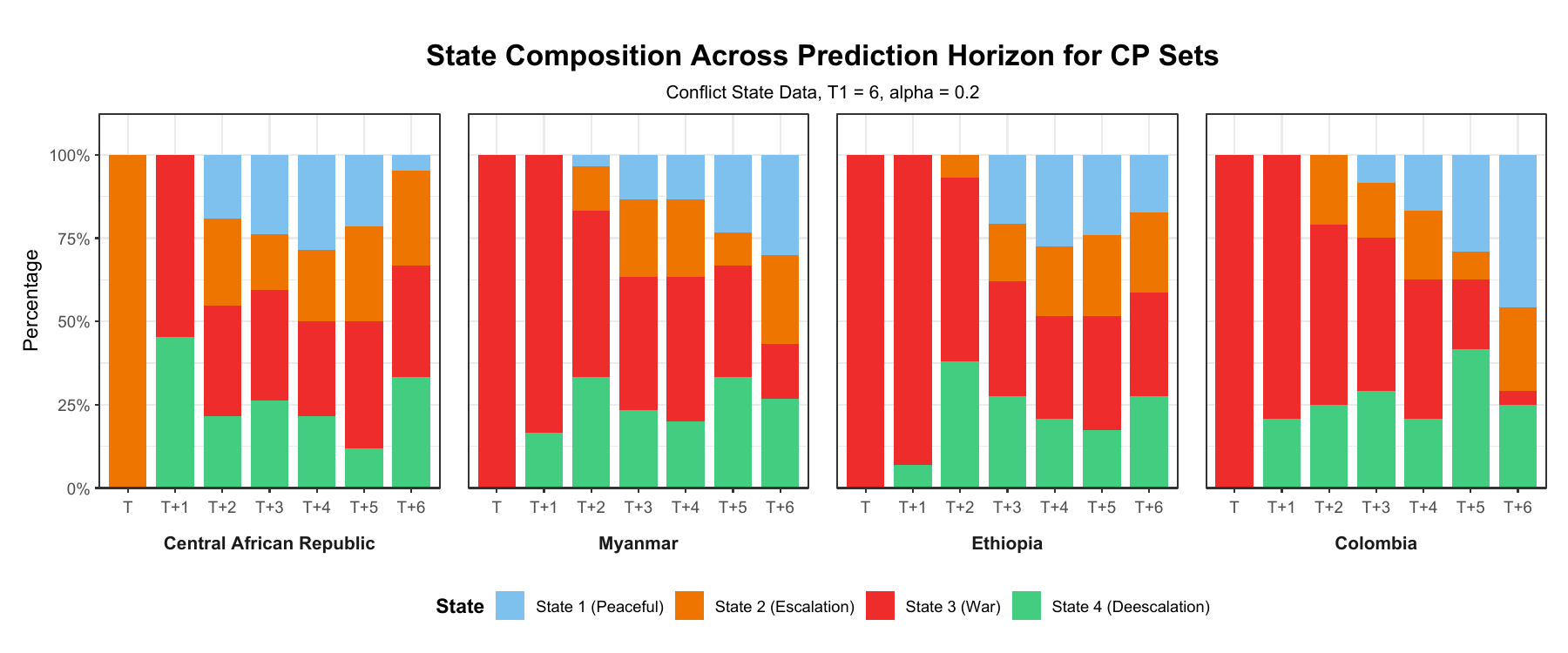}
    \includegraphics[width=0.7\textwidth, trim={0cm 0.5cm 0cm 0.5cm},clip]{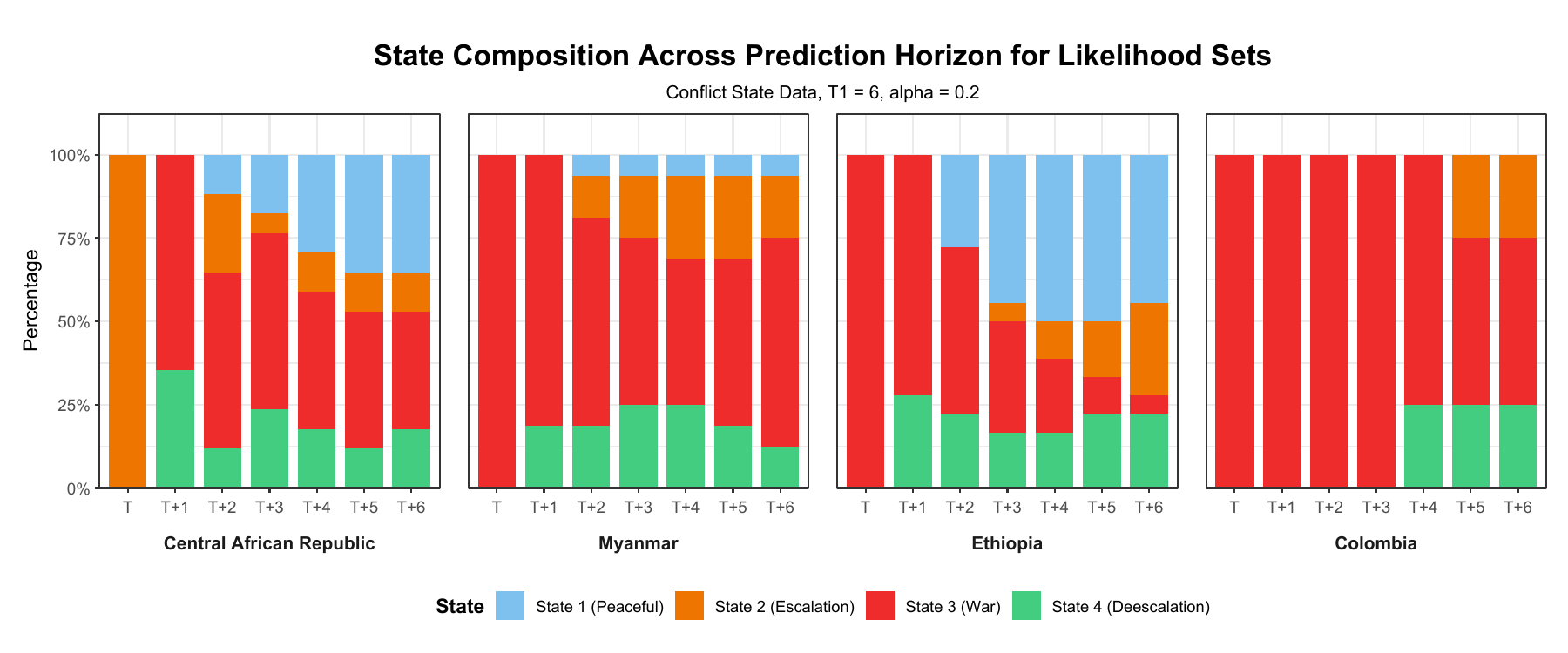}
    \caption{\footnotesize Composition of the prediction sets of forecasted state-sequences using CP (top row) and likelihood-based prediction (bottom row), where $T_1 = 6$ and $1-\alpha = 0.80$.}
    \label{fig:fatal_cp_set}
\end{figure}
The results in Figure \ref{fig:fatal_cp_set} highlight several important differences between the CP and likelihood-based approaches. While the CP and likelihood-based prediction sets are closely aligned over short forecast horizons, they diverge as the horizon extends. Notably, the likelihood-based sets converge quickly to a stationary distribution, effectively collapsing to a fixed, marginal state distribution as the time horizon increases. The CP sets, by contrast, retain greater coverage of plausible state-sequences across all forecast steps, reflecting the underlying trajectory uncertainty rather than steady-state behavior. In practical terms, this means CP sets are better positioned to capture the true dynamics of the forecasted sequence when projecting many steps ahead. For short forecast horizons this distinction is minor, but it becomes increasingly consequential when extending beyond the 6-month ($T_1 = 6$) horizon used here toward the 12- or 36-month horizons commonly targeted in the conflict forecasting literature. These patterns exhibited by the likelihood-based approach are visible for all four countries, but are particularly pronounced for Ethiopia and the Central African Republic which have observed conflict at time $T$ but have relatively low likelihood of observing conflict at time $T+6$. In contrast, the CP sets for these countries include a more diverse set of state-sequences that allow for the possibility of continued conflict at time $T+6$, which is more consistent with the observed conflict dynamics in these countries. In order to see more clearly the convergence of the likelihood-based approach to its stationary distribution, we present in Figure \ref{fig:predSet12} the compositions of both the conformal and likelihood-based prediction sets for $T_1 = 12$ (i.e., forecasting one year into the future) for specifically Ethiopia and the Central African Republic. Consistent with our reasoning before, the likelihood-based approach exhibits a clear convergent trend whereas the CP approach allows for a wider range of possible states in the later time points. We see stark differences between the two prediction approaches particularly in the expected conflict state at time $T+12$; the likelihood-based approach indicates a high likelihood of peace in both countries, whereas the CP approach indicates almost the complete reverse (i.e., high chance of conflict). Overall, these results highlight the importance of accounting for model misspecification and uncertainty in transition probabilities when forecasting conflict dynamics, particularly over longer time horizons.

\section{Relevant Edge-Case} \label{sec:limitations}
\begin{figure}[!htb]
    \centering
    \includegraphics[width=\linewidth, trim={0cm 0.5cm 0cm 0.5cm},clip]{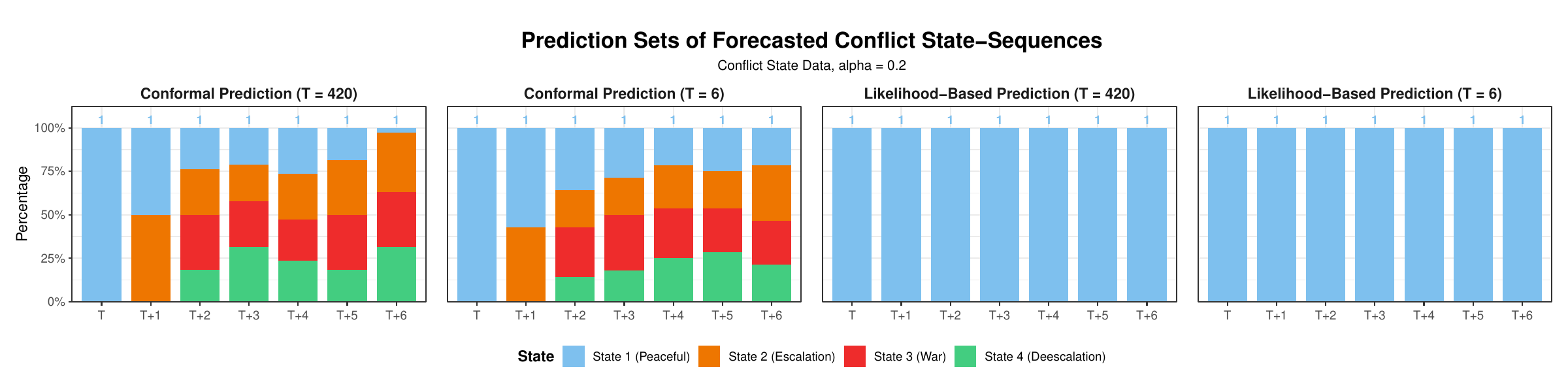}
    \caption{\footnotesize Composition of prediction sets for the special case where all of the calibration data are state 1. These data correspond to Sweden.}
    \label{fig:predSetSpecialCase}
\end{figure}

Recall the data cleaning procedure from Section \ref{sec:data_clean}. In particular, we exclude countries where a single state perpetuates for the entire duration of the observed time frame. What happens to the results when we \textit{do} include such countries? Consider Sweden, for example, where the country is in peacetime (state 1) for every observation. If we apply Algorithm \ref{alg:cpAlg} for $T_1 = 6$, Figure \ref{fig:predSetSpecialCase} illustrates the prediction sets from using both conformal and likelihood-based prediction.
Immediately, there is a stark difference between the CP and likelihood-based prediction approaches, moreso than any difference observed in any country in Figure \ref{fig:fatal_cp_set}. The CP results seem counter-intuitive because in the training/calibration data for Sweden, we have $T=420$ sequential months of state 1 suggesting that the prediction set would have high confidence in predicting a perpetuation of state 1; however, we do not see this. In fact, for $T=420$, the only forecasted state-sequence in the prediction set that ends with $X_{T+T_1} = 1$ is the state-sequence with $X_{T+1} = \hdots = X_{T+T_1} = 1$. Thus, why does the CP set contain such a large proportion of forecasted state-sequences that exhibit state-transitions uncharacteristic of the calibration sequence?

\begin{figure}[!htb]
\centering
\begin{tikzpicture}[font=\ttfamily, every node/.style={inner sep=2pt},scale=0.9]
%% =======================
%% Top Left Panel
%% =======================
\begin{scope}
    \node at (0,1.0) {\textbf{Case 1} $(i = 1)$};

    \node (a1) at (-3,0) {1};
    \node (a2) at (-2.175,0) {$\hdots$};
    \node (a3) at (-1.35,0) {1};
    \node (a4) at (-0.525,0) {\scriptsize$X_{T+1}$};
    \node (a5) at (0.3,0) {\scriptsize$X_{T+2}$};
    \node (a6) at (1.125,0) {\scriptsize$X_{T+3}$};
    \node (a7) at (1.95,0) {\scriptsize$X_{T+4}$};
    \node (a8) at (2.775,0) {\scriptsize$X_{T+5}$};
    \node (a9) at (3.6,0) {1};

    \draw [decorate,decoration={brace,amplitude=5pt,mirror}]
        ($(a1.south west)+(0,-0.2)$) -- ($(a1.south east)+(0,-0.2)$)
        node[midway,below=6pt] {$I_1$};
    \draw [decorate,decoration={brace,amplitude=5pt,mirror}]
        ($(a3.south west)+(0,-0.2)$) -- ($(a3.south east)+(0,-0.2)$)
        node[midway,below=6pt] {$I_T$};
    \draw[rounded corners] (-3.8,1.3) rectangle (4.4,-1.2);
\end{scope}
%% =======================
%% Top Right Panel
%% =======================
\begin{scope}[shift={(8.75,0)}]
    \node at (0,1.0) {\textbf{Case 2} $(i = 2)$};
    
    \node (a1) at (-3,0) {1};
    \node (a2) at (-2.175,0) {$\hdots$};
    \node (a3) at (-1.35,0) {1};
    \node (a4) at (-0.525,0) {\scriptsize$X_{T+1}$};
    \node (a5) at (0.3,0) {\scriptsize$X_{T+2}$};
    \node (a6) at (1.125,0) {\scriptsize$X_{T+3}$};
    \node (a7) at (1.95,0) {\scriptsize$X_{T+4}$};
    \node (a8) at (2.775,0) {\scriptsize$X_{T+5}$};
    \node (a9) at (3.6,0) {2};

    \draw [decorate,decoration={brace,amplitude=5pt,mirror}]
        ($(a1.south west)+(0,-0.2)$) -- ($(a3.south east)+(0,-0.2)$)
        node[midway,below=6pt] {$I_0$};
    \draw[rounded corners] (-3.8,1.3) rectangle (4.4,-1.2);
\end{scope}
%% =======================
%% Bottom Left Panel
%% =======================
\begin{scope}[shift={(0,-2.75)}]
    \node at (0,1.0) {\textbf{Case 3} $(i = 3)$};
    
    \node (a1) at (-3,0) {1};
    \node (a2) at (-2.175,0) {$\hdots$};
    \node (a3) at (-1.35,0) {1};
    \node (a4) at (-0.525,0) {\scriptsize$X_{T+1}$};
    \node (a5) at (0.3,0) {\scriptsize$X_{T+2}$};
    \node (a6) at (1.125,0) {\scriptsize$X_{T+3}$};
    \node (a7) at (1.95,0) {\scriptsize$X_{T+4}$};
    \node (a8) at (2.775,0) {\scriptsize$X_{T+5}$};
    \node (a9) at (3.6,0) {3};

    \draw [decorate,decoration={brace,amplitude=5pt,mirror}]
        ($(a1.south west)+(0,-0.2)$) -- ($(a3.south east)+(0,-0.2)$)
        node[midway,below=6pt] {$I_0$};
    \draw[rounded corners] (-3.8,1.3) rectangle (4.4,-1.2);
\end{scope}
%% =======================
%% Bottom Right Panel
%% =======================
\begin{scope}[shift={(8.75,-2.75)}]
    \node at (0,1.0) {\textbf{Case 4} $(i = 4)$};
    
    \node (a1) at (-3,0) {1};
    \node (a2) at (-2.175,0) {$\hdots$};
    \node (a3) at (-1.35,0) {1};
    \node (a4) at (-0.525,0) {\scriptsize$X_{T+1}$};
    \node (a5) at (0.3,0) {\scriptsize$X_{T+2}$};
    \node (a6) at (1.125,0) {\scriptsize$X_{T+3}$};
    \node (a7) at (1.95,0) {\scriptsize$X_{T+4}$};
    \node (a8) at (2.775,0) {\scriptsize$X_{T+5}$};
    \node (a9) at (3.6,0) {4};

    \draw [decorate,decoration={brace,amplitude=5pt,mirror}]
        ($(a1.south west)+(0,-0.2)$) -- ($(a3.south east)+(0,-0.2)$)
        node[midway,below=6pt] {$I_0$};
    \draw[rounded corners] (-3.8,1.3) rectangle (4.4,-1.2);
\end{scope}
\end{tikzpicture}
\caption{\footnotesize Illustration of how the number of permutable $i$-blocks changes depending on the last state, in the special case that all of the training/calibration data are the same state and $T_1 = 6$. Note that depending on the realizations for $(X_{T+1}, \hdots, X_{T+5})$, it is possible for $I_0$ to extend to further time points (i.e., beyond time $T$) for cases 2, 3, and 4.}
\label{fig:fourCaseSpecial}
\end{figure}
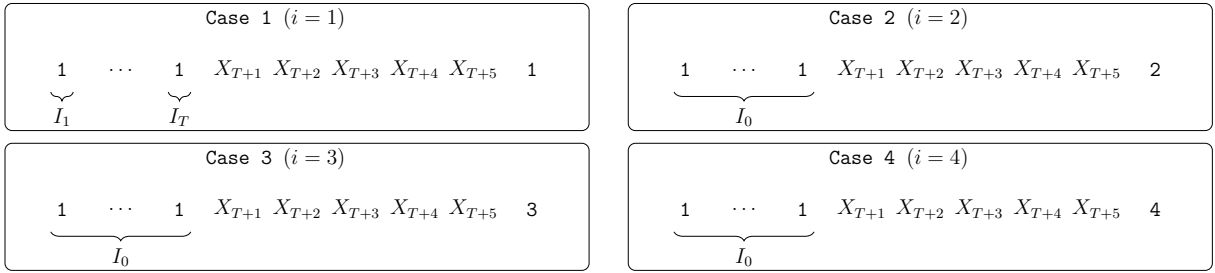

Recall that for each possible forecasted state-sequence, the CP algorithm splits the augmented sequence into permutable $i$-blocks, and then determines if the proposed forecasted sequence is included in the prediction set by comparing the nonconformity measure of the un-permuted augmented sequence, $S(\vb{I})$, to the nonconformity measures of all $i$-block permutations, $S(\pi)$ for $\pi \in \Pi$ (steps 5 and 6 of Algorithm \ref{alg:cpAlg}). The $i$-blocks are completely determined by the state of the last forecasted time point (i.e., $i=X_{T+T_1}$) and the number of $i$-blocks affects the number of possible permutations (i.e., $\qty|\Pi|$). Figure \ref{fig:fourCaseSpecial} provides an illustration of how the number of $i$-blocks changes depending on the value of $i = X_{T+T_1}$ in this special case where $X_1 = \hdots = X_T = 1$. 
Notice that when $i=1$, each time point in the calibration sequence is an individual $i$-block. Therefore, if $X_{T+1} \neq 1, \hdots$, or $X_{T+T_1-1} \neq 1$, $S(\vb{I})$ will be large whereas $S(\pi)$ will be small for almost all $\pi \in \Pi$. In other words, when $X_1 = \hdots = X_T = 1$ and $i=1$, the number of $i$-blocks (and therefore the number of possible permutations) is large enough to reject the inclusion of state-sequences in the prediction set that are dissimilar to the state-transitions exhibited in the calibration data (i.e., all state 1).

Contrarily, when $i\in \{2,3,4\}$, the number of $i$-blocks dramatically decreases, which has a profound effect on the sequences we accept into the prediction set. Specifically, the entire calibration sequence $\{X_t\}_{t=1}^T$ is contained within a single non-permutable block, $I_0$, as illustrated in Figure \ref{fig:fourCaseSpecial}. Therefore, for each possible forecasted state-sequence $(x_{T+1},...,x_{T+T_1}) \in \mathcal{X}^{T_1}$ with $x_{T+T_1} \neq 1$, $S(\vb{I})$ and $S(\pi)$ are no longer so disparate. In these cases, many of the proposed state-sequences are accepted when $X_{T+T_1} = x_{T+T_1} \neq 1$. If we consider the special case when $\qty|\Pi| = 1$ meaning $S(\vb{I}) = S(\pi)$, we see from step 6 of Algorithm \ref{alg:cpAlg} that the proposed state-sequence is accepted as an element of the prediction set with probability $1-\alpha$. The information-loss experienced in constructing CP sets when $i\in \{2,3,4\}$ and $X_1 = \hdots = X_T = 1$ can be further illustrated if we consider a smaller calibration size. Figure \ref{fig:predSetSpecialCase} also shows the composition of CP sets when the calibration length is only $T=6$. We see that the CP sets are similar when $T = 420$ and when $T = 6$ because, as Figure \ref{fig:fourCaseSpecial} illustrates, all of the calibration data gets grouped into a single non-permutable block, $I_0$, effectively reducing the calibration data to a single data point. A similar phenomenon can be seen in Figure \ref{fig:predSetSpecialCase} for the likelihood-based approach where the prediction sets for $T=420$ and $T=6$ are not just similar, but rather are the \textit{same}. This is because when $X_1 = \hdots = X_T = 1$, the estimated probability of remaining in state 1 is one, thus the prediction set defaults to the singleton set containing all state 1, regardless of the history length and forecast length.

This edge-case when $X_1, \hdots, X_T$ are all the same state is not a rare event; in our real data, there exists 105 out of the 191 countries where more than 99\% of the observations are state 1 or there exists fewer than five realizations of non-peaceful states (i.e., states 2, 3, or 4). Discussion of this edge-case, while not found in \cite{nettasinghe2023extending}, highlights an important difference between CP and the likelihood-based prediction. Notably, although the CP sets are inefficient, the marginal validity result still holds. We demonstrate that the CP sets are still valid under the partial-exchangeability assumption \textit{and} avoid degenerate prediction sets like that of the likelihood-based approach.

This edge-case touches on a more fundamental idea/phenomenon in statistics, one that dates back to the late 1700s / early 1800s when David Hume posited, and Pierre-Simon Laplace studied, the probability of the sun rising  \citep{zabell1988buffon}. Given recorded history, the sun has risen everyday, thus making the maximum likelihood estimate of the probability that the sun will rise equal to one. This poses a philosophical [rather than mathematical] conundrum because it eliminates \textit{uncertainty} for predicting future events (much like the results we observe from the likelihood-based predictions in Figure \ref{fig:predSetSpecialCase}). Laplace, instead, introduced a uniform prior distribution on the probability of the sun rising which leads to an estimated success probability of $(T+1)/(T+2)$, where $T$ is the number of historical observations. This prior avoids the certainty obtained using maximum likelihood estimation but introduces assumptions whose philosophical justification remains debated; more recently, \cite{martin2026induction} revisit these issues and develop a possibilistic inferential-model solution. The challenge of forecasting conflict state-sequences where all of the calibration data are from the same state is analogous to this problem of forecasting whether the sun will rise. Thus, for this edge-case, while CP provides larger prediction sets, we guarantee validity even under model misspecification; alternatively, the likelihood-based approach leads to the most efficient prediction sets, but at the cost of no uncertainty, regardless of the number of observations in the calibration data. Developing a more efficient CP algorithm that is robust to this edge-case is left for future work. Appendix \ref{app:artificial_endpoint} suggests a naive approach that illustrates the feasibility of constructing a more efficient CP algorithm; however, a validity guarantee is not provided nor is the interpretation of the prediction sets one-to-one to the CP approach described in Section \ref{subsec:cp}.

\section{Conclusion} \label{sec:conclusion}
Forecasting the conflict dynamics for a given country many months into the future has influence over a country's geopolitics. While a single point-prediction of a conflict state-sequence is easily interpretable, of tantamount importance is how certain that prediction is. Therefore, we present an application of CP for Markov processes in constructing prediction sets that are finite-sample valid for any prespecified confidence level. The utility of using this machine learning architecture is its robustness to model misspecification. Beyond achieving appropriate nominal coverage rates, a key advantage of the CP approach over likelihood-based simulation becomes increasingly apparent as the forecast horizon extends. Likelihood-based prediction sets may converge to a stationary distribution (collapsing to fixed state probabilities), thereby losing the ability to reflect genuine trajectory uncertainty. CP sets, by contrast, retain coverage of plausible state-sequences across all forecasted steps, preserving the temporal dynamics that are of substantive interest in many applications. 

For the purposes of this article, no covariate information is used in either prediction algorithms. That said, both the CP and likelihood-based prediction strategies naturally extend to the scenario where the transition probabilities between states depend on covariates.  We leave this to future work.

\bibliography{references}

\clearpage
\appendix

\section{Classical and Weighted CP Benchmarking} \label{app:addResults}

To assess the importance of accounting for Markovian dependence, we present two baseline CP procedures: split CP (``classical''), and recency-weighted CP (``weighted''). Algorithms \ref{alg:naiveCP} and \ref{alg:weightedCP} detail each framework adapted for a discrete-state Markov process. 

Algorithm \ref{alg:naiveCP} follows from the split CP framework described in Sections 1 and 1.1 of \cite{angelopoulos2021gentle}. Specifically, we partition the observed state-sequence into training and calibration sets, then estimate marginal state probabilities from the training data, and construct nonconformity scores using disjoint windows of length $T_1$ from the calibration data. Candidate forecasted sequences are then evaluated against the empirical distribution of calibration scores to construct prediction sets. This procedure does not explicitly account for the Markov dependency, thus the exchangeability assumption required for the finite-sample coverage guarantee does not generally hold.

Algorithm \ref{alg:weightedCP} extends the split CP framework by assigning greater importance to recent observations in the calibration data. This algorithm is based on the recency-weighting approach discussed in Section 4.6 of \cite{angelopoulos2021gentle}. Specifically, we introduce an exponential decay factor, $\lambda\in(0,1]$, to discount older calibration windows. While this weighting scheme places greater ``value'' on more recent observations, it does not explicitly account for the Markov assumption.

For both the classical and weighted CP algorithms, we split the data into 70\% training and 30\% calibration. For weighted CP, we set $\lambda = 0.98$ because it down-weights less recent calibration windows faster than the value used in Section 4.6 of \cite{angelopoulos2021gentle}.

\begin{figure}[!htbp]
    \centering
    \includegraphics[width=0.35\linewidth, trim={0cm 0.5cm 0cm 0.9cm},clip]{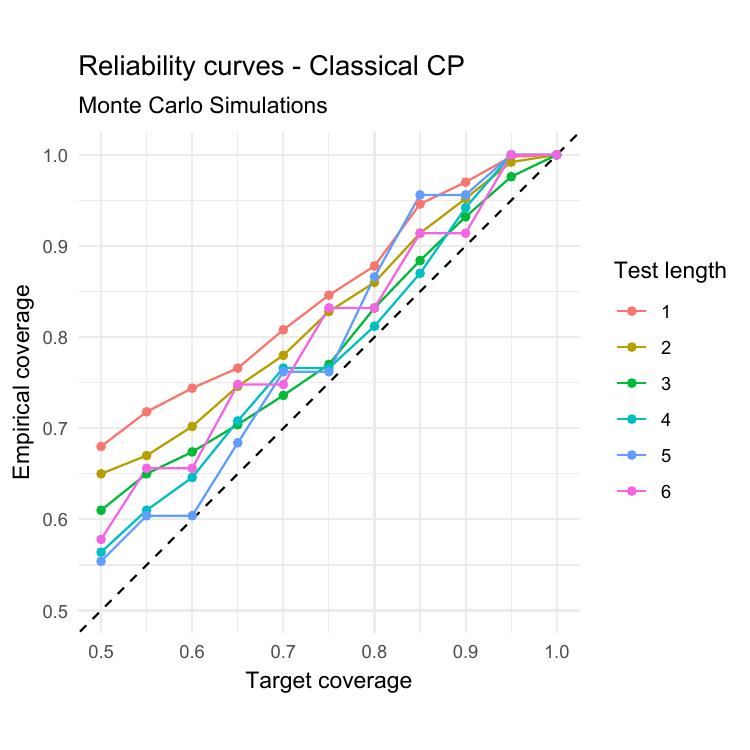}
    \includegraphics[width=0.35\linewidth, trim={0cm 0.5cm 0cm 0.9cm},clip]{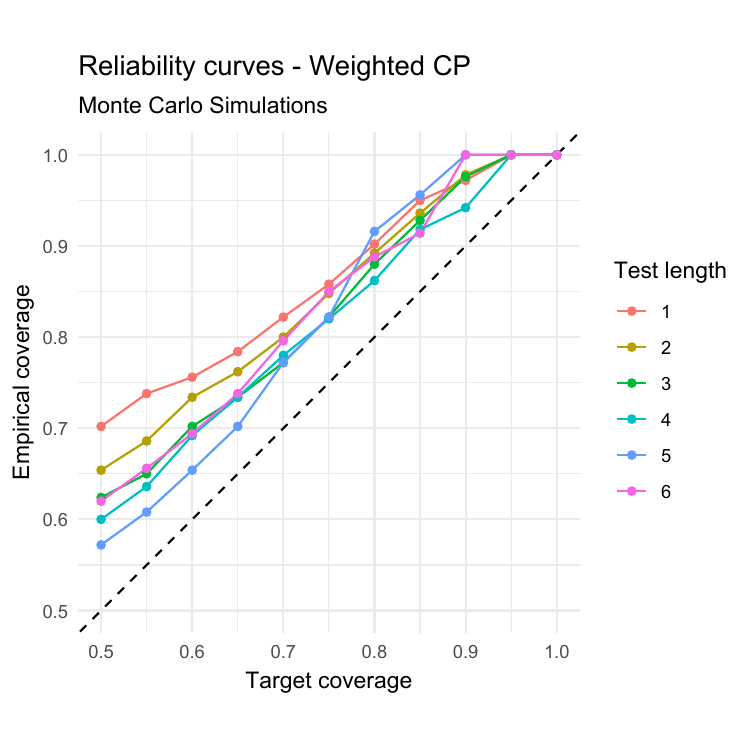}
    \caption{\footnotesize Reliability curves illustrating the relationship between the empirical coverage and the target coverage for both the classical CP approach (left) and weighted CP approach (right) for the simulated data.}
    \label{fig:sim_rel_curves_naive_weighted_cp}
\end{figure}

\begin{figure}[!htbp]
    \centering
    \includegraphics[width=0.7\linewidth, trim={0cm 0.5cm 0cm 0.5cm},clip]{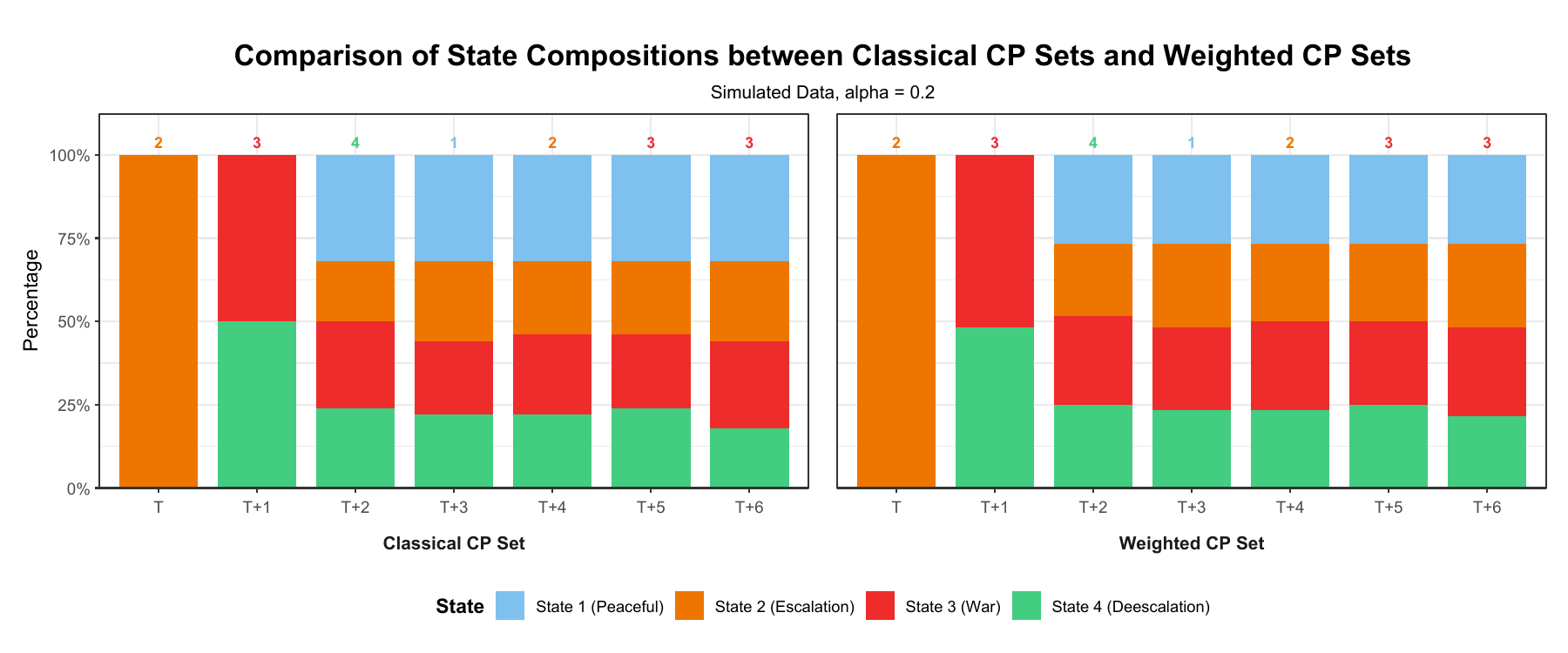}
    \caption{\footnotesize Visualization of the prediction sets for a single simulated state-sequence resulting from the classical CP approach (left) and weighted CP approach (right) for the simulated data.}
    \label{fig:sim_state_comp_naive_weighted_cp}
\end{figure}

We apply Algorithms \ref{alg:naiveCP} and \ref{alg:weightedCP} to the same simulated data that is described in Section \ref{sec:sim}. Subsequently, Figures \ref{fig:sim_rel_curves_naive_weighted_cp} and \ref{fig:sim_state_comp_naive_weighted_cp}, and Table \ref{tab:cardsets_naive_weighted} provide the analogous results. In particular, we see that both the classical and weighted CP algorithms tend to over-cover across most confidence levels and forecast horizons. Figure \ref{fig:sim_rel_curves_naive_weighted_cp} shows that this over-coverage is particularly pronounced at lower target coverage levels and shorter test lengths. In addition, although the weighted CP approach assigns greater importance to more recent observations, this modification does not lead to a substantial change in coverage relative to the classical CP approach. In comparison, as shown in the left panel of Figure \ref{fig:rel_curves_mc}, the partial-exchangeability CP approach produces well calibrated prediction sets across all test lengths and target coverage values.

Figure \ref{fig:sim_state_comp_naive_weighted_cp} provides a visualization for the state compositions of the CP sets produced by both the classical and weighted CP approaches, where $T$ = 200, $T_1$ = 6, and $1-\alpha$ = 0.80. Both approaches produce prediction sets with similar state compositions, particularly at the earlier forecasted time points. However, as the forecast horizon increases, the compositions of both prediction sets become relatively stable. Neither approach exhibits substantial variation in state composition across the later forecasted time points. Comparing these to the partial-exchangeability CP method, we see from Figure \ref{fig:set_vis_mc} that properly accounting for the Markov assumption leads to more meaningful and variable state compositions across the forecasted time points.

Lastly, Table \ref{tab:cardsets_naive_weighted} displays the cardinality of prediction sets produced by both the classical and weighted CP methods for $T_1 = 6$ and across 5 different target coverages. Both methods produce prediction sets of comparable sizes, although weighted CP produces slightly larger sets. As the test lengths and target coverages increase, the cardinalities of both methods increase substantially. For example, when \(T_1=6\) and \(1-\alpha=0.95\), both approaches include all 64 allowable future state-sequences in their prediction sets. In comparison, the partial-exchangeability CP approach from Section 4.2 produces prediction sets with an average cardinality of 50.19 under the same conditions. Combined with Figures \ref{fig:sim_rel_curves_naive_weighted_cp} and \ref{fig:sim_state_comp_naive_weighted_cp}, these results suggest that applying traditional CP methods to temporally dependent data can lead to unnecessarily large prediction sets without improving calibration, offering further motivation for utilizing the partial-exchangeability framework in Algorithm \ref{alg:cpAlg}.

\begin{table}[!htb]
\vspace{-0.1in}
\centering
\scriptsize
\begin{tabular}{c*{5}{rr}}
& \multicolumn{10}{c}{\textbf{Target Coverage}} \\
\cmidrule(lr){2-11}
& \multicolumn{2}{c}{0.55}
& \multicolumn{2}{c}{0.65}
& \multicolumn{2}{c}{0.75}
& \multicolumn{2}{c}{0.85}
& \multicolumn{2}{c}{0.95} \\
\cmidrule(lr){2-11}
\phantom{Horizon}
& Classical & Weighted
& Classical & Weighted
& Classical & Weighted
& Classical & Weighted
& Classical & Weighted \\
\midrule

$(T_1 = 1)$ & 0.82 & 0.88 & 0.93 & 0.97 & 1.27 & 1.30 & 1.68 & 1.74 & 2.00 & 2.00 \\
$(T_1 = 2)$ & 1.26 & 1.28 & 1.62 & 1.57 & 2.16 & 2.19 & 3.03 & 3.14 & 3.92 & 4.00 \\
$(T_1 = 3)$ & 2.07 & 2.26 & 3.10 & 3.42 & 3.86 & 4.45 & 5.55 & 6.24 & 6.99 & 8.00 \\
$(T_1 = 4)$ & 2.96 & 3.75 & 6.17 & 6.52 & 7.08 & 8.77 & 10.59 & 11.99 & 16.00 & 16.00 \\
$(T_1 = 5)$ & 7.93 & 7.56 & 11.96 & 12.66 & 12.95 & 15.81 & 24.66 & 24.66 & 32.00 & 32.00 \\
$(T_1 = 6)$ & 14.05 & 13.16 & 21.91 & 22.51 & 33.15 & 36.08 & 47.33 & 47.33 & 64.00 & 64.00
\end{tabular}
\caption{\footnotesize Comparison of the cardinality of prediction sets using classical CP and weighted CP for the simulation over different prediction lengths, $T_1$, and target coverage levels. The results are averaged over 100 simulations.}
\label{tab:cardsets_naive_weighted}
\end{table}

\section{Calibration-Guarantee for Alternative Likelihood-Based Approach}\label{app:likeProof}
The following proves that the {\em expected} cumulative probability mass from $\mathcal{C}^{\text{like}^*}_{1-\alpha}$ of (\ref{eq:cLikeStar}) will equal $1-\alpha$. For $K>1$,
{\scriptsize \begin{align*}
    &\mathbb{E}\qty{\qty[\sum_{j = 1}^K \mathbb{P}\qty(\vb*{X}_{T+1}^{T+T_1} = \vb*{x}^{(j)} \mid X_T=x_T)]\cdot \mathbf{1}\{U^* \leq p\} + \qty[\sum_{j = 1}^{K-1} \mathbb{P}\qty(\vb*{X}_{T+1}^{T+T_1} = \vb*{x}^{(j)} \mid X_T=x_T)]\cdot \mathbf{1}\{U^* > p\}}\\
    &\hspace{1in} = \qty[\sum_{j = 1}^K \mathbb{P}\qty(\vb*{X}_{T+1}^{T+T_1} = \vb*{x}^{(j)} \mid X_T=x_T)]\cdot \mathbb{P}\{U^* \leq p\} + \qty[\sum_{j = 1}^{K-1} \mathbb{P}\qty(\vb*{X}_{T+1}^{T+T_1} = \vb*{x}^{(j)} \mid X_T=x_T)]\cdot \mathbb{P}\{U^* > p\}\\
    &\hspace{1in} = \qty[\sum_{j = 1}^K \mathbb{P}\qty(\vb*{X}_{T+1}^{T+T_1} = \vb*{x}^{(j)} \mid X_T=x_T)]\cdot p + \qty[\sum_{j = 1}^{K-1} \mathbb{P}\qty(\vb*{X}_{T+1}^{T+T_1} = \vb*{x}^{(j)} \mid X_T=x_T)]\cdot (1-p)\\
    &\hspace{1in} = p \cdot \mathbb{P}\qty(\vb*{X}_{T+1}^{T+T_1} = \vb*{x}^{(K)} \mid X_T=x_T) + \sum_{j = 1}^{K-1} \mathbb{P}\qty(\vb*{X}_{T+1}^{T+T_1} = \vb*{x}^{(j)} \mid X_T=x_T)\\
    &\hspace{1in} = (1-\alpha) - \sum_{j = 1}^{K-1} \mathbb{P}\qty(\vb*{X}_{T+1}^{T+T_1} = \vb*{x}^{(j)} \mid X_T=x_T) + \sum_{j = 1}^{K-1} \mathbb{P}\qty(\vb*{X}_{T+1}^{T+T_1} = \vb*{x}^{(j)} \mid X_T=x_T)\\
    &\hspace{1in} = 1-\alpha.
\end{align*}}%

\section{Naive ``Solution'' for Efficient CP}\label{app:artificial_endpoint}

\begin{figure}[!htbp]
    \centering
    \includegraphics[width=0.6\linewidth, trim={0cm 0.5cm 0cm 0.5cm},clip]{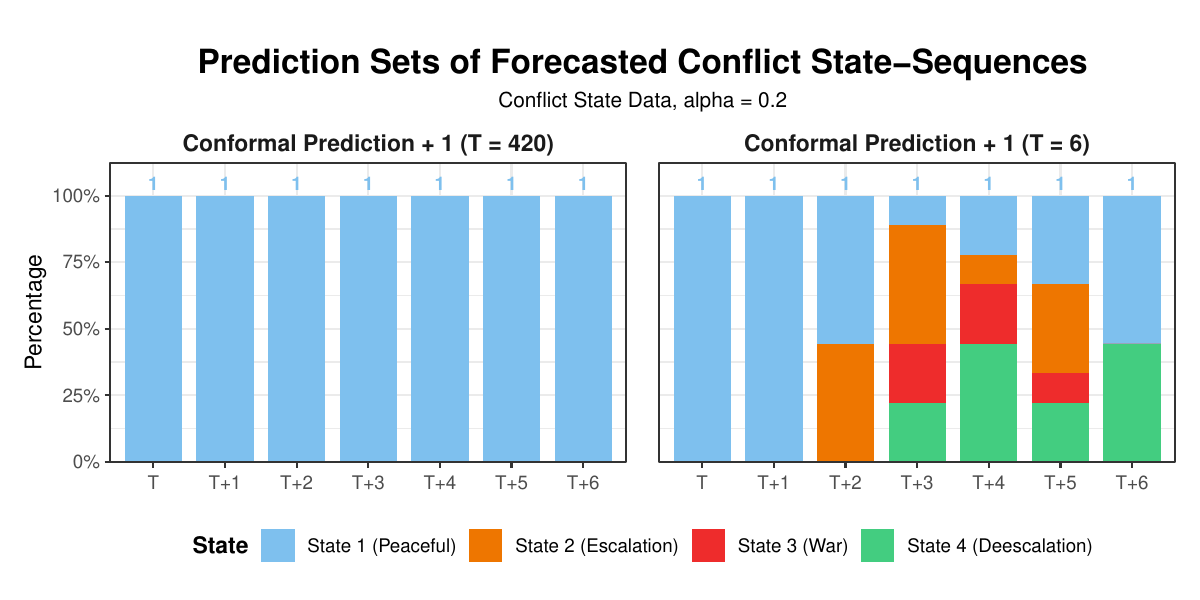}
    \caption{\footnotesize Composition of prediction sets for the special case where all of the calibration data are state 1. These data correspond to Sweden.}
    \label{fig:app:naive}
\end{figure} 

We present a naive solution to the efficiency loss seen in Figure \ref{fig:predSetSpecialCase} by introducing an artificial $T+T_1 + 1$ time point where $X_{T+T_1+1} = 1$ always. Figure \ref{fig:app:naive} presents the resulting CP sets under the title ``Conformal Prediction + 1''. We see in the left column of Figure \ref{fig:app:naive} with $T=420$ that this new CP approach is as efficient as the likelihood-based approach where the prediction set contains one state-sequence, namely the sequence $X_{T+1} = \hdots = X_{T+T_1} = 1$. Then, when $T=6$, this new CP approach produces a prediction set which includes more than one state-sequence (9 sequences to be exact) which accounts for the added uncertainty from the smaller calibration length. The reason we say this is a ``naive'' solution is because our interpretation of the results now slightly differs. Before adding this new $X_{T+T_1+1} = 1$ time point, we could interpret the prediction sets in Figure \ref{fig:predSetSpecialCase} as the uncertainty quantification for a predicted state-sequence $T_1$ time points into the future; however, with the additional constraint of $X_{T+T_1+1} = 1$, the interpretation of the prediction sets in Figure \ref{fig:app:naive} now becomes the uncertainty quantification for a predicted state-sequence $T_1$ time points into the future, \textit{assuming the country returns to peacetime at time $T+T_1+1$}. It is also important to note that the validity guarantee for the CP algorithm in Section \ref{subsec:cp} need not apply for this conditional CP method; hence, we cannot guarantee the validity of this naive solution. We leave the construction of a more efficient, valid CP method to handle this edge-case as future work, but highlight this edge-case as an important discovery that was not discussed/addressed in \cite{nettasinghe2023extending}.

\vspace{1in}
\begin{algorithm}[!htbp]
    \scriptsize

    \caption{Classical CP for discrete-state sequences}
    \label{alg:naiveCP}

    \begin{algorithmic}
        \renewcommand{\baselinestretch}{0.86}\selectfont

        % --------------------------------------------------
        % Input and Output
        % --------------------------------------------------

        \State \parbox[t]{0.95\linewidth}{%
            \textbf{Input}: Calibration data $\{X_t\}_{t=1}^T$,
            miscoverage $\alpha \in [0,1]$,
            prediction length $T_1$,
            fitting fraction $\gamma \in (0,1)$,
            and adjacency matrix $\vb{A}$.
        }

        \State \parbox[t]{0.95\linewidth}{%
            \textbf{Output}: Prediction set
            $\mathcal{C}_{1-\alpha} \subseteq \mathcal{X}^{T_1}$,
            where $\mathcal{X}^{T_1}$ is the set of all possible
            state-sequences of length $T_1$.
        }

        \Statex \vspace{0.4em}

        % --------------------------------------------------
        % Step 1: Split historical data
        % --------------------------------------------------

        \State 1. \parbox[t]{0.95\linewidth}{%
            Split $\{X_t\}_{t=1}^T$ into fitting and calibration sets,
            where $T_{\mathrm{fit}}$ is the number of fitting observations,
            $\mathcal{D}_{\mathrm{fit}}$ is the fitting set, and
            $\mathcal{D}_{\mathrm{cal}}$ is the calibration set:
        }

        \State \hspace{1em}$
            \displaystyle
            \begin{aligned}
                T_{\mathrm{fit}} &= \lfloor \gamma T \rfloor,\\
                \mathcal{D}_{\mathrm{fit}}
                    &= \{X_t\}_{t=1}^{T_{\mathrm{fit}}},\\
                \mathcal{D}_{\mathrm{cal}}
                    &= \{X_t\}_{t=T_{\mathrm{fit}}+1}^{T}.
            \end{aligned}
        $

        \Statex \vspace{0.4em}

        % --------------------------------------------------
        % Step 2: Estimate marginal probabilities
        % --------------------------------------------------

        \State 2. \parbox[t]{0.95\linewidth}{%
            Estimate the marginal state probabilities
            from the fitting set using Laplace smoothing:
        }

        \State \hspace{1em}$
            \displaystyle
            \widehat{p}(s) =
            \frac{
                1 + \sum_{t=1}^{T_{\mathrm{fit}}}
                \mathbf{1}(X_t=s)
            }{
                T_{\mathrm{fit}} + |\mathcal{X}|
            },
            \quad s \in \mathcal{X}.
        $

        \Statex \vspace{0.4em}

        % --------------------------------------------------
        % Step 3: Construct calibration windows
        % --------------------------------------------------

        \State 3. \parbox[t]{0.95\linewidth}{%
            Construct disjoint calibration windows
            $\vb*{z}_r =
            (X_{T_{\mathrm{fit}}+(r-1)T_1+1},
            \ldots,X_{T_{\mathrm{fit}}+rT_1})$,
            for $r=1,\ldots,N$, where
            $N=\lfloor (T-T_{\mathrm{fit}})/T_1 \rfloor$.
            Discard any remaining incomplete window.
        }

        \Statex \vspace{0.4em}

        % --------------------------------------------------
        % Step 4: Calibration scores
        % --------------------------------------------------

        \State 4. Calculate the nonconformity score
        for each calibration window:

        \State \hspace{1em}$
            \displaystyle
            S(\vb*{z}_r) =
            -\sum_{j=0}^{T_1-1}
            \log \widehat{p}
            (X_{T_{\mathrm{fit}}+(r-1)T_1+1+j}).
        $

        \Statex \vspace{0.4em}

        % --------------------------------------------------
        % Candidate evaluation
        % --------------------------------------------------

        \For{$\vb*{x}=(x_{T+1},\ldots,x_{T+T_1})
            \in \mathcal{X}^{T_1}$}

            \Statex \vspace{0.2em}

            % Step 5: Feasibility

            \State 5. \parbox[t]{0.88\linewidth}{%
                If $\vb*{x}$ violates an allowable transition
                in $\vb{A}$, set $\widehat{q}(\vb*{x})=0$
                and continue to the next candidate.
            }

            \Statex \vspace{0.3em}

            % Step 6: Candidate score

            \State 6. Calculate the nonconformity score
            $S(\vb*{x})$:

            \State \hspace{1em}$
                \displaystyle
                S(\vb*{x}) =
                -\sum_{j=1}^{T_1}
                \log \widehat{p}(x_{T+j}).
            $

            \Statex \vspace{0.3em}

            % Step 7: P-value

            \State 7. Calculate the p-value:

            \State \hspace{1em}$
                \displaystyle
                \widehat{q}(\vb*{x}) =
                \frac{
                    1 +
                    \sum_{r=1}^{N}
                    \mathbf{1}\{
                        S(\vb*{z}_r) \geq S(\vb*{x})
                    \}
                }{
                    N+1
                }.
            $

        \EndFor

        \Statex \vspace{0.4em}

        % --------------------------------------------------
        % Output
        % --------------------------------------------------

        \State \textbf{return}
        $\mathcal{C}_{1-\alpha} =
        \{\vb*{x} \in \mathcal{X}^{T_1} :
        \widehat{q}(\vb*{x}) > \alpha\}$

    \end{algorithmic}
\end{algorithm}

\begin{algorithm}[!htbp]
    \scriptsize

    \caption{Recency-weighted CP for discrete-state sequences}
    \label{alg:weightedCP}

    \begin{algorithmic}
        \renewcommand{\baselinestretch}{0.86}\selectfont

        % --------------------------------------------------
        % Input and Output
        % --------------------------------------------------

        \State \parbox[t]{0.95\linewidth}{%
            \textbf{Input}: Calibration data $\{X_t\}_{t=1}^T$,
            miscoverage $\alpha \in [0,1]$,
            prediction length $T_1$,
            fitting fraction $\gamma \in (0,1)$,
            decay factor $\lambda \in (0,1]$,
            and adjacency matrix $\vb{A}$.
        }

        \State \parbox[t]{0.95\linewidth}{%
            \textbf{Output}: Prediction set
            $\mathcal{C}_{1-\alpha} \subseteq \mathcal{X}^{T_1}$,
            where $\mathcal{X}^{T_1}$ is the set of all possible
            state-sequences of length $T_1$.
        }

        \Statex \vspace{0.4em}

        % --------------------------------------------------
        % Step 1: Split historical data
        % --------------------------------------------------

        \State 1. \parbox[t]{0.95\linewidth}{%
            Split $\{X_t\}_{t=1}^T$ into fitting and calibration sets,
            where $T_{\mathrm{fit}}$ is the number of fitting observations,
            $\mathcal{D}_{\mathrm{fit}}$ is the fitting set, and
            $\mathcal{D}_{\mathrm{cal}}$ is the calibration set:
        }

        \State \hspace{1em}$
            \displaystyle
            \begin{aligned}
                T_{\mathrm{fit}} &= \lfloor \gamma T \rfloor,\\
                \mathcal{D}_{\mathrm{fit}}
                    &= \{X_t\}_{t=1}^{T_{\mathrm{fit}}},\\
                \mathcal{D}_{\mathrm{cal}}
                    &= \{X_t\}_{t=T_{\mathrm{fit}}+1}^{T}.
            \end{aligned}
        $

        \Statex \vspace{0.4em}

        % --------------------------------------------------
        % Step 2: Estimate marginal probabilities
        % --------------------------------------------------

        \State 2. \parbox[t]{0.95\linewidth}{%
            Estimate the marginal state probabilities
            from the fitting set using Laplace smoothing:
        }

        \State \hspace{1em}$
            \displaystyle
            \widehat{p}(s) =
            \frac{
                1 + \sum_{t=1}^{T_{\mathrm{fit}}}
                \mathbf{1}(X_t=s)
            }{
                T_{\mathrm{fit}} + |\mathcal{X}|
            },
            \quad s \in \mathcal{X}.
        $

        \Statex \vspace{0.4em}

        % --------------------------------------------------
        % Step 3: Construct calibration windows
        % --------------------------------------------------

        \State 3. \parbox[t]{0.95\linewidth}{%
            Construct disjoint calibration windows
            $\vb*{z}_r =
            (X_{T_{\mathrm{fit}}+(r-1)T_1+1},
            \ldots,X_{T_{\mathrm{fit}}+rT_1})$,
            for $r=1,\ldots,N$, where
            $N=\lfloor (T-T_{\mathrm{fit}})/T_1 \rfloor$.
            Discard any remaining incomplete window.
        }

        \Statex \vspace{0.4em}

        % --------------------------------------------------
        % Step 4: Calibration scores
        % --------------------------------------------------

        \State 4. Calculate the nonconformity score
        for each calibration window:

        \State \hspace{1em}$
            \displaystyle
            S(\vb*{z}_r) =
            -\sum_{j=0}^{T_1-1}
            \log \widehat{p}
            (X_{T_{\mathrm{fit}}+(r-1)T_1+1+j}).
        $

        \Statex \vspace{0.4em}

        % --------------------------------------------------
        % Step 5: Recency weights
        % --------------------------------------------------

        \State 5. \parbox[t]{0.95\linewidth}{%
            Assign each calibration window an
            exponentially decaying weight based
            on its recency:
        }

        \State \hspace{1em}$
            \displaystyle
            \begin{aligned}
                w_r &= \lambda^{T-(T_{\mathrm{fit}}+rT_1)},\\
                W &= \sum_{r=1}^{N}w_r.
            \end{aligned}
        $

        \Statex \vspace{0.4em}

        % --------------------------------------------------
        % Candidate evaluation
        % --------------------------------------------------

        \For{$\vb*{x}=(x_{T+1},\ldots,x_{T+T_1})
            \in \mathcal{X}^{T_1}$}

            \Statex \vspace{0.2em}

            % Step 6: Feasibility

            \State 6. \parbox[t]{0.88\linewidth}{%
                If $\vb*{x}$ violates an allowable transition
                in $\vb{A}$, set $\widehat{q}(\vb*{x})=0$
                and continue to the next candidate.
            }

            \Statex \vspace{0.3em}

            % Step 7: Candidate score

            \State 7. Calculate the nonconformity score
            $S(\vb*{x})$:

            \State \hspace{1em}$
                \displaystyle
                S(\vb*{x}) =
                -\sum_{j=1}^{T_1}
                \log \widehat{p}(x_{T+j}).
            $

            \Statex \vspace{0.3em}

            % Step 8: Weighted p-value

            \State 8. Calculate the weighted p-value:

            \State \hspace{1em}$
                \displaystyle
                \widehat{q}(\vb*{x}) =
                \frac{
                    1 +
                    \sum_{r=1}^{N}
                    w_r
                    \mathbf{1}\{
                        S(\vb*{z}_r) \geq S(\vb*{x})
                    \}
                }{
                    W+1
                }.
            $

        \EndFor

        \Statex \vspace{0.4em}

        % --------------------------------------------------
        % Output
        % --------------------------------------------------

        \State \textbf{return}
        $\mathcal{C}_{1-\alpha} =
        \{\vb*{x} \in \mathcal{X}^{T_1} :
        \widehat{q}(\vb*{x}) > \alpha\}$

    \end{algorithmic}
\end{algorithm}

\end{document}